\documentclass[12pt,preprint]{aastex}

\newcommand{\bdv}[1]{\mbox{\boldmath$#1$}}

\def\au{{\rm AU}} 
 
\def\kms{{\rm km}\,{\rm s}^{-1}}
\def\masyr{{\rm mas}\,{\rm yr}^{-1}}
\def\kpc{{\rm kpc}}

\def\rel{{\rm rel}}

\def\hel{{\rm hel}}
\def\geo{{\rm geo}}
\def\e{{\rm E}}
\def\bpi{{\bdv\pi}}
\def\bmu{{\bdv\mu}}

\def\bv{{\bf v}}

\usepackage{subfloat}
\begin{document}
\title{Pathway to the Galactic Distribution of Planets: Combined
{\it Spitzer} and Ground-Based
Microlens Parallax Measurements of 21 Single-Lens Events}

\author
{
S.~Calchi~Novati$^{1,2,3,}$\altaffilmark{*},
A.~Gould$^{4,U}$,
A.~Udalski$^{5,O}$,
J.~W.~Menzies$^{6,P}$,
I.~A.~Bond$^{7,M}$,
Y.~Shvartzvald$^{8,W}$,
R.~A.~Street$^{9,R,S}$,
M.~Hundertmark$^{10,11,S,R}$,
C.~A.~Beichman$^1$,
J.~C.~Yee$^{12,U,}$\altaffilmark{@},
S.~Carey$^{13}$
\vspace{-0.3cm}
\begin{center}
and
\end{center}
\vspace{-0.3cm}
R.~Poleski$^{5,4}$,
J.~Skowron$^{5}$,
S.~Koz{\l}owski$^{5}$,
P.~Mr{\'o}z$^{5}$,
P.~Pietrukowicz$^{5}$,
G.~Pietrzy{\'n}ski$^{5,14}$
M.~K.~Szyma{\'n}ski$^{5}$,
I.~Soszy{\'n}ski$^{5}$,
K.~Ulaczyk$^{5}$,
{\L}.~Wyrzykowski$^{5,15}$
\vspace{-0.3cm}
\begin{center}
(The OGLE collaboration)
\end{center}
\vspace{-0.3cm}
M.~Albrow$^{16}$,
J.~P.~Beaulieu$^{17,18}$,
J.~A.~.R.~Caldwell$^{19}$,
A.~Cassan$^{17}$,
C.~Coutures$^{17}$,
C.~Danielski$^{20}$,
D.~Dominis~Prester$^{21}$,
J.~Donatowicz$^{22}$,
K.~Lon\v{c}ari\'{c}$^{21}$,
A.~McDougall$^{16}$,
J.~C.~Morales$^{18}$,
C.~Ranc$^{17}$,
W.~Zhu$^{4}$
\vspace{-0.3cm}
\begin{center}
(The PLANET collaboration)
\end{center}
\vspace{-0.3cm}
F.~Abe$^{23}$,
R.~K.~Barry$^{24}$,
D.~P.~Bennett$^{25}$,
A.~Bhattacharya$^{25}$,
D.~Fukunaga$^{23}$,
K.~Inayama$^{26}$, 
N.~Koshimoto$^{27}$,
S.~Namba$^{27}$,
T.~Sumi$^{27}$, 
D.~Suzuki$^{25}$,
P.~J.~Tristram$^{28}$,
Y.~Wakiyama$^{23}$,
A.~Yonehara$^{26}$
\vspace{-0.3cm}
\begin{center}
(The MOA collaboration)
\end{center}
\vspace{-0.3cm}
D.~Maoz$^{8}$,
S.~Kaspi$^{8}$,
M.~Friedmann$^{8}$
\vspace{-0.3cm}
\begin{center}
(The Wise group)
\end{center}
\vspace{-0.3cm}
E.~Bachelet$^{29}$,
R.~Figuera~Jaimes$^{11,30,S}$,
D.~M.~Bramich$^{29,S}$,
Y.~Tsapras$^{9,31,S}$,
K.~Horne$^{11}$,
C.~Snodgrass$^{32,33,S}$,
J.~Wambsganss$^{34,S}$,
I.~A.~Steele$^{35}$,
N.~Kains$^{30}$
\vspace{-0.3cm}
\begin{center}
(The RoboNet collaboration)
\end{center}
\vspace{-0.3cm}
V.~Bozza$^{2,36}$,
M.~Dominik$^{11,R,}$\altaffilmark{\#},
U.~G.~J{\o}rgensen$^{10}$,
K.~A.~Alsubai$^{29}$,
S.~Ciceri$^{37}$,
G.~D'Ago$^{2,36,R}$,
T.~Haugb{\o}lle$^{10}$,
F.~V.~Hessman$^{38}$,
T.~C.~Hinse$^{39}$,
D.~Juncher$^{10}$,
H.~Korhonen$^{10,40}$,
L.~Mancini$^{37}$,
A.~Popovas$^{10}$,
M.~Rabus$^{37,41}$,
S.~Rahvar$^{42}$,
G.~Scarpetta$^{2,36,3}$,
R.~W.~Schmidt$^{34,R}$,
J.~Skottfelt$^{10}$,
J.~Southworth$^{43}$,
D.~Starkey$^{11}$,
J.~Surdej$^{44}$,
O.~Wertz$^{44}$,
M.~Zarucki$^{3}$
\vspace{-0.3cm}
\begin{center}
(The MiNDSTEp consortium)
\end{center}
\vspace{-0.3cm}
B.~S.~Gaudi$^{4}$,
R.~W.~Pogge$^{4}$,
D.~L.~DePoy$^{45}$
\vspace{-0.3cm}
\begin{center}
(The $\mu$FUN collaboration)
\end{center}
\vspace{-0.3cm}
\normalsize{$^{1}$NASA Exoplanet Science Institute, MS 100-22, California Institute of Technology, Pasadena, CA 91125, USA} \\
\normalsize{$^{2}$Dipartimento di Fisica ``E. R. Caianiello'', Universit\`a di Salerno, Via Giovanni Paolo II, 84084 Fisciano (SA),\ Italy} \\
\normalsize{$^{3}$Istituto Internazionale per gli Alti Studi Scientifici (IIASS), Via G. Pellegrino 19, 84019 Vietri Sul Mare (SA), Italy}\\
\normalsize{$^{4}$Department of Astronomy, Ohio State University, 140 W. 18th Ave., Columbus, OH  43210, USA} \\
\normalsize{$^{5}$Warsaw University Observatory, Al.~Ujazdowskie~4, 00-478~Warszawa, Poland} \\ 
\normalsize{$^{6}$South African Astronomical Observatory, PO Box 9, Observatory 7935, South Africa} \\
\normalsize{$^{7}$Institute of Natural and Mathematical Sciences, Massey University, Private Bag 102-904, North Shore Mail Centre, Auckland, New Zealand}\\
\normalsize{$^{8}$School of Physics and Astronomy, Tel-Aviv University, Tel-Aviv 69978, Israel} \\
\normalsize{$^{9}$Las Cumbres Observatory Global Telescope Network, 6740 Cortona Drive, suite 102, Goleta, CA 93117, USA}\\
\normalsize{$^{10}$Niels Bohr Institutet, K{\o}benhavns Universitet, Juliane Maries Vej 30, 2100 K{\o}benhavn {\O}, Denmark} \\
\normalsize{$^{11}$SUPA, School of Physics \& Astronomy, North Haugh, University of St Andrews, KY16 9SS, Scotland, UK} \\
\normalsize{$^{12}$Harvard-Smithsonian Center for Astrophysics, 60 Garden St., Cambridge, MA 02138, USA} \\
\normalsize{$^{13}$Spitzer Science Center, MS 220-6, California Institute of Technology,Pasadena, CA, USA} \\
\normalsize{$^{14}$Universidad de Concepci{\'o}n, Departamento de Astronomia, Casilla 160--C, Concepci{\'o}n, Chile} \\
\normalsize{$^{15}$Institute of Astronomy, University of Cambridge, Madingley Road, Cambridge CB3 0HA, UK} \\
\normalsize{$^{16}$University of Canterbury, Dept. of Physics and Astronomy, Private Bag 4800, 8020 Christchurch, New Zealand} \\
\normalsize{$^{17}$Institut d'Astrophysique de Paris, UPMC-CNRS, UMR 7095, 98bis Boulevard Arago, 75014 Paris, France} \\
\normalsize{$^{18}$LESIA, Observatoire de Paris, Section de Meudon 5, place Jules Janssen 92195 Meudon, France } \\
\normalsize{$^{19}$McDonald Observatory, 16120 St Hwy Spur 78 \#2, Fort Davis, TX 79734, USA} \\
\normalsize{$^{20}$Institut d'Astrophysique Spatiale,  UMR 8617, Universit\'e Paris Sud 91405 Orsay, France} \\
\normalsize{$^{21}$Department of Physics, University of Rijeka, Radmile Matej v{c}i\'{c} 2, 51000 Rijeka, Croatia} \\
\normalsize{$^{22}$Technical University of Vienna, Department of Computing,Wiedner Hauptstrasse 10, 1040 Wien, Austria} \\
\normalsize{$^{23}$Solar-Terrestrial Environment Laboratory, Nagoya University,
Nagoya 464-8601, Japan}\\
\normalsize{$^{24}$Astrophysics Science Division, NASA Goddard Space Flight Center, Greenbelt, MD 20771, USA}\\
\normalsize{$^{25}$Department of Physics, University of Notre Dame, Notre Dame, IN 46556, USA}\\
\normalsize{$^{26}$Department of Physics, Faculty of Science, Kyoto Sangyo University, 603-8555 Kyoto, Japan}\\
\normalsize{$^{27}$Department of Earth and Space Science, Graduate School of Science, Osaka University, Toyonaka, Osaka 560-0043, Japan}\\
\normalsize{$^{28}$Mt. John University Observatory, P.O. Box 56, Lake Tekapo 8770, New Zealand}\\
\normalsize{$^{29}$Qatar Environment and Energy Research Institute, Qatar Foundation, P.O. Box 5825, Doha, Qatar}\\
\normalsize{$^{30}$European Southern Observatory, Karl-Schwarzschild-Str. 2, 85748 Garching bei M\"unchen, Germany}\\
\normalsize{$^{31}$School of Physics and Astronomy, Queen Mary University of London, Mile End Road, London E1 4NS, UK}\\
\normalsize{$^{32}$The Open University, Walton Hall, Milton Keynes MK7 6AA, UK} \\
\normalsize{$^{33}$Max Planck Institute for Solar System Research, Justus-von-Liebig-Weg 3, 37077 G\"ottingen, Germany}\\
\normalsize{$^{34}$Astronomisches Rechen-Institut, Zentrum f{\"u}r Astronomie der Universit{\"a}t Heidelberg (ZAH), M\"onchhofstra{\ss}e 12-14, 69120 Heidelberg, Germany} \\
\normalsize{$^{35}$Astrophysics Research Institute, Liverpool John Moores University, Liverpool CH41 1LD, UK}\\
\normalsize{$^{36}$Istituto Nazionale di Fisica Nucleare, Sezione di Napoli, Via Cintia, 80126 Napoli, Italy}\\
\normalsize{$^{37}$Max Planck Institute for Astronomy, K\"onigstuhl 17, 69117 Heidelberg, Germany} \\
\normalsize{$^{38}$Institut f\"ur Astrophysik, Georg-August-Universit\"at G\"ottingen, Friedrich-Hund-Platz 1, 37077 G\"ottingen, Germany} \\
\normalsize{$^{39}$Korea Astronomy and Space Science Institute, 776 Daedeokdae-ro, Yuseong-gu, 305-348 Daejeon, Republic of Korea} \\
\normalsize{$^{40}$Finnish Centre for Astronomy with ESO (FINCA), University of Turku, V{\"a}is{\"a}l{\"a}ntie 20, FI-21500 Piikki{\"o}, Finland} \\
\normalsize{$^{41}$Instituto de Astrof\'isica, Facultad de F\'isica, Pontificia Universidad Cat\'olica de Chile, Av.\ Vicu\~na Mackenna 4860,  7820436 Macul, Santiago, Chile} \\
\normalsize{$^{42}$Department of Physics, Sharif University of Technology, P.\,O.\,Box 11155-9161 Tehran, Iran} \\
\normalsize{$^{43}$Astrophysics Group, Keele University, Staffordshire, ST5 5BG, UK} \\
\normalsize{$^{44}$Institut d'Astrophysique et de G\'eophysique,  Universit\'e de Li\`ege, 4000 Li\`ege, Belgium} \\
\normalsize{$^{45}$Department of Physics and Astronomy, Texas A\& M University, College Station, TX 77843-4242, USA} 
\vspace{-0.3cm}
\begin{center}
\normalsize{$^{O}$(The OGLE collaboration)}\\
\normalsize{$^{P}$(The PLANET collaboration)}\\
\normalsize{$^{M}$(The MOA collaboration)}\\
\normalsize{$^{W}$(The Wise group)}\\
\normalsize{$^{R}$(The RoboNet collaboration)}\\
\normalsize{$^{S}$(The MiNDSTEp consortium)}\\
\normalsize{$^{U}$(The $\mu$FUN collaboration)}\\
\end{center}
\vspace{-0.3cm}
}
\altaffiltext{*}{Sagan Visiting Fellow.}
\altaffiltext{@}{Sagan Fellow.}
\altaffiltext{\#}{Royal Society University Research Fellow.}


\begin{abstract}

We present microlens parallax measurements for 21 (apparently) isolated
lenses observed toward the Galactic bulge that were imaged simultaneously 
from Earth and {\it Spitzer},
which was $\sim 1\,\au$ West of Earth in projection.  We combine
these measurements with a kinematic model of the Galaxy to derive
distance estimates for each lens, with error bars that are small 
compared to the Sun's Galactocentric distance.  The ensemble therefore
yields a well-defined cumulative distribution of lens distances. 
 In principle it is possible to compare this
distribution against a set of planets detected in the same experiment
in order to measure the Galactic distribution of planets.
Since these {\it Spitzer} observations yielded
only one planet, this is not yet possible in practice.  However,
it will become possible as larger samples are accumulated.

\end{abstract}

\keywords{gravitational lensing: micro --- planets}

\section{{Introduction}
\label{sec:intro}}

It has been known for 50 years \citep{liebes64,refsdal64}
that microlensing measurements are
plagued by a severe degeneracy between the lens mass $M$, the source-lens
relative parallax $\pi_\rel= \au(D_L^{-1}-D_S^{-1})$, 
and the geocentric lens-source relative proper motion $\bmu_\geo$
(\citealt{gaudi12}, Equations (1) and (17)),
\begin{equation}
t_\e = {\theta_\e\over\mu_\geo};
\qquad
\theta_\e^2\equiv \kappa M \pi_\rel;
\qquad
\kappa\equiv {4 G\over c^2\au}\simeq 8.14{{\rm mas}\over M_\odot}.
\label{eqn:tedef}
\end{equation}
Here $\theta_\e$ is the angular Einstein radius and $t_\e$ is
the Einstein-radius crossing time in the ground-based reference frame.
It has also been known
for 50 years \citep{refsdal66} that the best way to systematically
ameliorate this degeneracy is to observe the events simultaneously
from solar orbit in order to measure the microlens parallax vector $\bpi_\e$,
\begin{equation}
\bpi_\e \equiv {\pi_\rel\over \theta_\e}\,{\bmu\over\mu}\,,
\label{eqn:piedef}
\end{equation}
where $\bmu$ can be the lens-source relative proper motion
in either the geocentric or heliocentric frame, in which cases $\bpi_\e$
is the representation in the same frame.
(Note that, as the ratio of two angles, $\bpi_\e$ is dimensionless.) 
If $\bpi_\e$ is measured one obtains
strong constraints on $M$ and $\pi_\rel$ from
\begin{equation}
M = {\theta_\e\over \kappa\pi_\e} = {\mu_\geo t_\e\over \kappa\pi_\e},
\qquad
\pi_\rel = \theta_\e\pi_\e = \mu_\geo t_\e \pi_\e.
\label{eqn:partial}
\end{equation}
Hence, even if $\theta_\e$ is not measured (as it almost never is for
single-lens microlenses),  $M$ and $\pi_\rel$ can be estimated
fairly robustly just from the fact that the great majority of the
microlenses have $\mu_\geo$ within a factor 2 of $\mu_\geo\sim 4\,\masyr$.
However, without the additional information from $\pi_\e$,
the three physical quantities $M$, $\pi_\rel$ and $\mu_\geo$ cannot
be disentangled from the single measured parameter $t_\e$, so
that, for example, $M$ remains uncertain by an order of magnitude
\citep{gould00}.  

Nevertheless, before 2014, there was only one space-based parallax
measurement out of more than 10,000 recorded microlensing
events: \citet{dong07} used {\it Spitzer} to measure the microlens
parallax of a rare (almost unique) bright event toward the 
Small Magellanic Cloud, OGLE-2005-SMC-001.

{\it Spitzer} \citep{spitzer04} 
has several advantages but also important disadvantages
as a possible ``microlens parallax satellite'' for observations
toward the Galactic bulge.  First, of course,
it is in solar orbit, gradually drifting behind Earth at somewhat more
than $0.1\,\au\,{\rm yr}^{-1}$.  Hence, by now it trails Earth by almost
$90^\circ$.  Second, at $3.6\,\mu$m, its IRAC camera \citep{irac04}
has relatively good
resolution of $\sim 2^{\prime\prime}$, not much worse than the resolutions of the
ground-based surveys that discover and monitor the events.  Third,
it can be pointed at targets on relatively short notice.  This is
important because microlensing events typically peak (and then decline)
within a few weeks of their discovery.  Hence, either the satellite
must be able to respond quickly \citep{refsdal66,gould94} or it must,
like ground observatories, survey an extended field in hope of
detecting events from previously unidentified sources \citep{gouldhorne}.

{\it Spitzer}'s most important disadvantage is that due to Sun-angle
viewing restrictions, it can observe
any given target that lies near the ecliptic (including the
entire Galactic bulge, which hosts $> 99\%$ of all recorded microlensing
events) for only two $\sim 38\,$day intervals per year.  Moreover, during
only one of these is it possible to simultaneously
observe the bulge from Earth (and
so measure parallaxes).  Second, while {\it Spitzer}'s real-time response
can in principle be extremely rapid, such rapid responses are also 
very disruptive to its overall mission and science return.  By contrast,
the normal (non-disruptive) response time to unexpected requests for data
is of order a month, which would be useless for the great majority
of microlensing events.  For completeness, we mention that 
the fact that {\it Spitzer}
observes at wavelengths far redward of those used for ground-based
microlensing observations was initially believed to be a problem
(e.g., \citealt{gould99}), but this now appears to be relatively 
minor (e.g., \citealt{ob140939}).

The observations reported here derive from a 100-hour ``pilot program''
awarded by the Director to demonstrate the
feasibility of {\it Spitzer} microlens parallax observations.
The scientific objectives were of course driven by the overall
potential of {\it Spitzer} to determine microlens masses, particularly for
planetary events.  However, these objectives were also sculpted by the
challenges discussed in the previous paragraph, and in particular by
the need to demonstrate concretely that these challenges could
be met.  For example, working with {\it Spitzer} operations, we
developed a new observing protocol for ``regular'' (non-disruptive)
target-of-opportunity observations with 3--9 day turnaround.
The times of
the microlensing observations were preplanned to occur in blocks
approximately once per 24 hours, but the targets for a given week 
were not finalized
until a few days before they were uploaded to the spacecraft
(Section~2 and Figure~1 of \citealt{ob140124} and also Section~\ref{sec:spitz}
of this paper).
In particular, the weekly 
observing decisions made under this protocol were aimed primarily
at maximizing the number of successful parallax measurements,
while making an extra effort to measure parallaxes for as many
binary and planetary events as possible.

An alternative strategy might have been to develop purely objective
criteria for the weekly choices of targets and cadences.  The
ensemble of parallax measurements of isolated lenses made using 
such an objective protocol could then be forward modeled to extract 
the underlying mass function, as envisaged by \citet{han95}.

The reasons for not using purely objective criteria 
were three-fold.  First, as stated above,
the overwhelming objective was to determine feasibility, which
can best be done by learning from successful measurements.  Second,
it is very difficult to develop objective criteria without concrete
experience (exactly the point of a ``pilot program'').  Finally,
the lens mass function is not the most important scientific result
that can be extracted from an ensemble of isolated-lens measurements.

Rather, the most critical application of an
ensemble of isolated-lens parallaxes is to serve as the
comparison sample by which the planet detections can be transformed
into a measurement of the Galactic distribution of planets.  That is,
as long as the planetary events are not chosen for {\it Spitzer}
observations because they are known to have planets, then the planetary
events can be considered to be ``drawn fairly'' from the ensemble of
(mainly) isolated-lens events, regardless of whether the process by
which the latter are chosen can be modeled or not.  This also
means that if, during successive years of observations, the
selection criteria change, the planet sample and the isolated-lens
sample can each be concatenated, and they will still yield a fair
comparison.  This situation is analogous to the selection
of high-magnification events for intensive followup that led to the
first microlens-planet frequency analysis \citep{gould10},  the most 
relevant point in both cases being that events are selected for 
observations without regard to whether or not they have planets in them.

Now, since there was only one planet\footnote{In fact OGLE-2014-BLG-0298,
which showed a perturbation that was strongly suspected to be planetary
in nature well before the commencement of {\it Spitzer} observations,
was aggressively monitored during this campaign.  However, exactly
because these observations were triggered by the (suspected) presence of
a planet, this event is not part of the ``fair sample'' and is therefore
not considered in the present work.  The value of these {\it Spitzer}
observations, as with {\it Spitzer} observations of known binary microlensing
events, is to measure the mass of a potentially interesting object,
rather than for statistical studies.  To date, planetary anomalies far out
on the rising wing of the lightcurve (like OGLE-2014-BLG-0298) have constituted
roughly 8\% of all planetary events, so elimination of these events from
the Galactic-distribution sample is not likely to be a major loss.
However, if future planet surveys have more uniform lightcurve 
coverage than past ones (e.g.,  \citealt{henderson14})
then this fraction will increase somewhat.}
in the {\it Spitzer} ``pilot program''
sample \citep{ob140124}, it is not yet possible to derive a Galactic
distribution of planets.  Nevertheless, it is important to make an
initial effort to measure the distance distribution of the isolated lens
sample, partly to learn practically how to do this from real data and
partly to understand what type of lenses were effectively selected
by the selection procedures used in the ``pilot program''.  Even
though these procedures cannot be comprehensibly quantified, they
do have quantifiable elements (like 3--9 day delay times) that
by themselves select for certain types of lenses.  Even a qualitative
understanding of these effects may influence the choice of objective
selection criteria in future years.  Thus, although it is clearly
too early to measure the Galactic distribution of planets, it is
actually quite urgent to begin those components of the analysis
that can be done.

Making a statistical estimate of the distance distribution of 
the ensemble of isolated lenses requires that a probability distribution 
be assigned to the distance of each lens.  In general, this probability
distribution will be much more compact if the well-known parallax
degeneracy \citep{refsdal66,gould94} is broken, as it was by
\citet{ob140939} for the case of OGLE-2014-BLG-0939.  That is, 
because $u$ (and so $u_0$) 
enters the lensing magnification equation quadratically
(Equation~(\ref{eqn:pac}) below), space-based
parallax measurements generically have a four-fold degeneracy 
in the vector microlens parallax $\bpi_\e$,
\begin{equation}
\bpi_\e = {\au\over D_\perp}\biggl({\Delta t_0\over t_\e},
\Delta u_{0,\pm,\pm}\biggr),
\label{eqn:sbpar}
\end{equation}
where the $x$-axis is defined by the direction of the projected
Earth-satellite separation vector ${\bf D}_\perp$, 
$\Delta t_0 = t_{0,\rm sat} - t_{0,\oplus}$ is the difference in times
of maximum as seen at the two locations, 
$\Delta u_{0,-,\pm} = \pm(|u_{0,\rm sat}| - |u_{0,\oplus}|)$ is the difference in impact
parameters assuming that they are on the same side of the lens, and
$\Delta u_{0,+,\pm} = \pm(|u_{0,\rm sat}| + |u_{0,\oplus}|)$ is their difference 
assuming that they are on the opposite sides.  While the two solutions
$\Delta u_{0,-,\pm}$ (or two solutions $\Delta u_{0,+,\pm}$) yield very similar
$\pi_\e$ (so $M$ and $\pi_\rel$), these two {\it sets of solutions} can have very
different $\pi_\e$ from each other.  The overall sign of $\Delta u_0$,
which is designated by the second ``$\pm$'' subscript, is positive if
$u_{0,\oplus}>0$, which by convention occurs if
``the lens passes the moving source on its right as seen from Earth''
\citep{gould04}. See his Figure~2.

\subsection{Rich Argument} \label{sec:rich}

The present {\it ensemble} of parallax measurements
provides the first opportunity to test an idea that to our knowledge
was first suggested by James Rich (private communication, ca.\ 1997),
but never (to our knowledge) written up. 
Rich's original idea was that the two components of $\bpi_\e D_\perp/\au$, 
(namely $\Delta\tau = \Delta t_0/t_\e$ and $\Delta u_0$) should
in general be of the same order.
This is true for different classes of lenses for different reasons.   
If the lens
is in the bulge, then the direction of relative proper motion $\bmu$
(and so $\bpi_\e$) is nearly randomly distributed over a circle.  Similarly,
if the lens is close to the Sun (i.e., within about 1 kpc) then the
direction of proper motion is primarily determined by the lens
peculiar motion and is again basically random.  Finally, if the lens
is at intermediate distances in the Galactic disk, then its proper motion
should be roughly aligned with Galactic rotation, which in ecliptic
coordinates (relevant since ${\bf D}_\perp$ is closely aligned with the ecliptic)
has comparable components. 

According to original Rich's idea,
in the case that the true
solution is one of $\Delta u_{0,-,\pm}$ solutions, the two components
will generally be roughly equal $|\Delta\tau|\sim |\Delta
u_{0,-,\pm}|$. If these components are themselves small, $|\Delta
u_{0,-,\pm}|\ll |u_{0,\oplus}|$, then the components for the other
solution will be highly unequal, $|\Delta u_{0,+,\pm}|\sim
2|u_{0,\oplus}|\gg |\Delta u_{0,-,\pm}|$ 
and consequently $|\Delta u_{0,+,\pm}| \gg |\Delta\tau|$.  Hence,
seeing such roughly equal components for one solution and highly
unequal components for the other, one should conclude that the first
solution is probably correct.

In the course of working on the events in this paper, we realized
that Rich's argument can be extended to apply constraints from two
degrees of freedom, rather than just one.  This increases the
argument's statistical
power considerably.  Properly speaking it should then be called the
``Extended Rich argument'', but for simplicity we continue to
simply say ``Rich argument''.  We begin by noting that the
parallax amplitude basically has a two-fold degeneracy, which we
denote $\pi_{\e,\pm}$, corresponding to\footnote{For simplicity 
of notation we will neglect 
the second $\pm$ in $\Delta u_{0,\pm,\pm}$ 
for the remainder of this section.}  
$(|\Delta u_{0,\pm}|,\Delta\tau)$.
One of these is the actual parallax
$\pi_{\e,\rm{true}}$ and the other is spurious, $\pi_{\e,\rm{false}}$.  However,
it is often the case that the lightcurve does not distinguish
between these.  Nevertheless, we can
define a theoretical quantity
\begin{equation}\label{eq:rich_eps}
\epsilon =  {\pi_{\e,\rm{false}}\over \pi_{\e,\rm{true}}}
=\biggl({
(|u_{0,\rm sat}|-|u_{0,\oplus}|)^2 + (\Delta\tau)^2\over
(|u_{0,\rm sat}|+|u_{0,\oplus}|)^2 + (\Delta\tau)^2}
\biggr)^{\pm 1/2}
= \biggl({
(|\Delta u_{0,-}|)^2 + (\Delta\tau)^2\over
(|\Delta u_{0,+}|)^2 + (\Delta\tau)^2}
\biggr)^{\pm 1/2}\,,
\end{equation}
where the sign refers to the cases $\pi_{\e,\rm{true}}= \pi_{\e,\pm}$.
Hence, for $\pi_{\e,\rm{true}}= \pi_{\e,+}$,
$\epsilon<1$, and if $\pi_{\e,+}\gg\pi_{\e,-}$, 
then $\epsilon\ll 1$.
We can test the hypothesis that $\pi_{\e,\rm{true}}= \pi_{\e,+} \gg \pi_{\e,-}$
by asking what is the probability of 
$\epsilon\leq\epsilon_0$ where $\epsilon_0\ll 1$.
This can be divided into two questions: first, what is the prior
probability of $\epsilon <1$ given that $\pi_\e$ has some given
true value, and second, what is the conditional probability
$\epsilon<\epsilon_0$ given that $\epsilon<1$.

The first probability (namely that $\pi_{\e,\rm{true}}=\pi_{\e,+}$) is
certainly less than unity, and typically of order one half.
We do not further investigate this probability because it
depends on the details of event selection and because its specific
value has marginal impact on the overall result.  

If $\epsilon\ll 1$, then $|\Delta u_{0,-}| \la \epsilon|\Delta
u_{0,+}|$ and $\Delta\tau \la \epsilon|\Delta u_{0,+}|$.  Under
this hypothesis, the latter condition gives highly unequal 
components for $\pi_{\e,\rm{true}} = \pi_{\e,+}$, implying a very special 
angle $\alpha$ for the lens-source relative
motion with respect to the direction perpendicular to the
Earth-satellite axis, $|\sin\alpha|<\epsilon$, whereas a priori,
$\alpha$ could assume any direction over the circle.  This is the
basis of the original Rich argument.  However, the first condition
also constrains $|\Delta u_{0,-}|$ to a very narrow interval relative to
the full range of possibilities $-|\Delta u_{0,+}| < |\Delta u_{0,-}| <
|\Delta u_{0,+}|$ over which this quantity would be expected to be
uniformly distributed.  Eliminating duplicate geometries, we should
evaluate the probability that 
$[\alpha^2 + (\Delta u_{0,-}/\Delta u_{0,+})^2]<\epsilon_0^2$ 
under conditions where $\alpha$ is
uniformly distributed over $[-\pi/2,\pi/2]$ and $(\Delta
u_{0,-}/\Delta u_{0,+})$ is uniformly distributed over $[-1,1]$.  This
probability is just
\begin{equation} \label{eq:rich_prob}
P(\epsilon<\epsilon_0 | \epsilon<1) = {\pi\epsilon_0^2\over 2\pi}
= {\epsilon_0^2\over 2}.
\end{equation}
That is, the probability of $\epsilon<\epsilon_0 \ll 1$ (which
requires $\pi_{\e,\rm{true}}= \pi_{\e,+}$) is very small.

We next note that if $\pi_{\e,\rm{true}}\ll 1$, then the probability
of $\epsilon\gg 1$ is of order unity.  This is because 
$\pi_{\e,\rm{true}}\ll 1$ requires that $|\Delta u_0|\ll 1$ (and
$\Delta\tau\ll 1$).  Hence, under typical conditions, i.e., $u_0\sim
{\cal O}(0.5)$, we have $\Delta u_{0,+}\sim 2 |u_0| \sim {\cal O}(1)
\gg \Delta u_{0,-}$ and similarly $\Delta u_{0,+}\gg \Delta\tau$.

Hence, if we find $\pi_{\e,+}\gg\pi_{\e,-}$, then it is highly likely that
the $\pi_{\e,-}$ solution is correct.  
This is because, if $\pi_{\e,-}$ is correct, 
then we naturally expect the alternate
solution ($\pi_{\e,+}$) to be much bigger (i.e., $|\Delta
u_{0,+}|\gg|\Delta u_{0,-}|$ and $|\Delta u_{0,+}|\gg\Delta\tau$).
However, if the $\pi_{\e,+}$ solution were correct, then
we would expect the alternate solution ($\pi_{\e,-}$)
to be of the same general order,
and, in particular, the chance that the alternate solution was
as small as observed or smaller would be $\epsilon^2/2$.

Such an argument cannot be considered decisive
in any particular case because the proper motion can by chance be very
nearly perpendicular to ${\bf D}_\perp$ and the values of $\Delta u_0$
as seen from Earth and the satellite can by chance happen
to have very nearly equal magnitudes but opposite signs.
Nevertheless, if the objective is to find the cumulative 
distribution of lens distances
(rather than to securely determine the distance to a particular lens) then
it is appropriate to give unequal-component solutions lower
statistical weight when combining the distance estimates of the ensemble
to form a cumulative distribution.  

{\section{Observations}
\label{sec:obs}}

{\subsection{OGLE Observations}
\label{sec:ogleobs}}

All 21 of the events analyzed in this paper were discovered by 
the Optical Gravitational Lens Experiment (OGLE) 
based on observations with the 1.4 deg$^2$ camera on its 
1.3m Warsaw Telescope at the Las Campanas Observatory in Chile
using its Early Warning System (EWS) real-time event detection
software \citep{ews1,ews2}.  The observations reported here are entirely
in $I$ band, although some $V$ observations were also taken with the
aim of determining the source color.  The specific role of such 
source-color  measurements in the present study is discussed in 
Section~\ref{sec:color}.


{\subsection{Spitzer Observations}
\label{sec:spitz}}

The structure of our {\it Spitzer} observing protocol is described
in detail in Section 3.1 of \citet{ob140124}.  In brief, observations were made
during 38 2.63 hr windows between HJD$^\prime\equiv$HJD$-2450000 = 6814.0$
and 6850.0.  Each observation consisted of six dithered 30s exposures
in a fixed pattern using the $3.6\,\mu$m channel on the IRAC camera.  
Observation
sequences were uploaded to {\it Spitzer} operations on Mondays at UT 15:00,
for observations to be carried out Thursday to Wednesday (with slight
variations).  As described in \citet{ob140124}, J.C.Y. and A.G. 
balanced various
criteria to determine which targets to observe and how often.  In general,
there were too many targets to be able to observe all viable targets 
during each epoch.  The relation between weekly ``decision dates'' and
subsequent observations is illustrated in Figure~\ref{fig:lc} 
of \citet{ob140124}.

With three exceptions, the OGLE alerts for all 21 events occurred prior
to the first ``decision date'' (June 2 UT 15:00, HJD$^\prime$ 6811.1).
The alerts for 
OGLE-2014-BLG-1021, 
OGLE-2014-BLG-1049, and
OGLE-2014-BLG-1147, were announced on June 4, 6, and 18, respectively.
Hence, the first two could be observed only during four weeks, while the
third could be observed only during the final two weeks.

Table~\ref{tab:evt_param} lists the equatorial coordinates, 
ecliptic latitude, and number
of {\it Spitzer} observations for each event.  The ecliptic latitude
is important because in the limit that both the event and the {\it Spitzer}
spacecraft were directly on the ecliptic, the directional degeneracies
$\Delta u_{0,-,\pm}$ and $\Delta u_{0,+,\pm}$ could not be broken, even in
principle \citep{ob03238,ob09020}.

\subsection{Additional Lightcurve Data}

Additional lightcurve data were obtained for a total of 15 of the 21
events reported here from a total of 13 telescopes.  The MOA collaboration
\citep{bond01,sumi13}
obtained data on seven events as part of their normal survey operations
using a broad R/I filter on their 1.8m telescope at Mt.~John, New Zealand.
Similarly the Wise Collaboration 
\citep{sm12} obtained survey data on five events
using an $I$ band filter on their 1.0m telescope at Mitzpe Ramon, Israel.

Four other teams specifically targeted the {\it Spitzer} sample
for followup observations, all in $I$ band (or SDSS $i$ band).
The PLANET collaboration \citep{planet} observed
six events using the 1.0m Elizabeth telescope at Sutherland, South Africa.
The RoboNet/LCOGT (Las Cumbres Observatory Global Telescope Network) 
collaboration \citep{tsapras09} observed a total of four events from
a total of eight 1.0m telescopes in CTIO, Chile, Sutherland, South
Africa, and Siding Spring, Australia.  The MiNDSTEp 
(Microlensing Network for the Detection of Small Terrestrial Exoplanets)
consortium \citep{dominik10} 
observed four events from their 1.54m telescope at ESO La Silla, Chile,
and four events using the 0.35m Salerno University
telescope in Salerno, Italy.

Of the 21 events (6,9,1,1,2,1,1) were observed by (1,2,3,4,9,10,11) 
telescopes, respectively. We refer to Table~\ref{tab:evt_param}
for full details on the additional data set used.

{\subsection{Additional Color Data}
\label{sec:color}}

The $\mu$FUN (Microlensing Follow Up Network)
collaboration obtained a very limited quantity
of data on each of the 21 events using the ANDICAM \citep{depoy03} dichroic
camera on the SMARTS-CTIO 1.3m telescope.  These observations were
made simultaneously in $I$ and $H$ band and were for the specific
purpose of inferring the $I-[3.6]$ source color using an
$(I-H)$ vs.\ $(I-[3.6])$ instrumental color-color diagram. \citet{mb11293}
demonstrated for the case of MOA-2011-BLG-293 that
this color-color method could reliably constrain the source flux even
if a given data set lacked sufficient coverage for an independent flux
determination from the model. The incorporation of this constraint is
discussed further in Section~\ref{sec:anal}.

At the time of the decision to acquire these
data, it was deemed especially important to acquire $H$-band data because
it was unknown whether the extrapolation from the (more routinely taken) $V/I$
data to $3.6\,\mu$m would be feasible.  In fact, in most cases, the
OGLE $V$-band data did prove adequate to determine the $(I-[3.6])_S$
source color, but in five cases the source was either too red to obtain
reliable $V$-band data or OGLE did not happen to observe the event
in $V$ band when it was sufficiently magnified to determine $V-I$.
In all but one of these cases (OGLE-2014-BLG-0337), the $H$ band data
could be used to determine the source color 
(OGLE-2014-BLG-0805, OGLE-2014-BLG-0866, OGLE-2014-BLG-0944, 
OGLE-2014-BLG-1021).

{\subsection{Reductions}
\label{sec:reductions}}

With one exception the {\it Spitzer} data were reduced using 
the photometry tools available within MOPEX, a
package designed to analyse IRAC data \citep{mopex05}:
the analysis has been carried out
with aperture photometry for 6 events
(OGLE-2014-BLG-0099, OGLE-2014-BLG-0337, OGLE-2014-BLG-0589, 
OGLE-2014-BLG-0805, OGLE-2014-BLG-0944 and OGLE-2014-BLG-1021) and,
to better deal with crowding, 
for all the remaining ones,
with a point source response functions (PRFs) 
based photometry\footnote{For a specific discussion 
of PRFs fitting in IRAC data we refer to the online manual for MOPEX
http://irsa.ipac.caltech.edu/data/SPITZER/docs/dataanalysistools/tools/mopex/mopexusersguide/.}.
The exception was OGLE-2014-BLG-1049, which was reduced using
DoPhot \citep{dophot}.
All other lightcurve data were reduced using image subtraction
\citep{alard98}.  The CTIO $H$-band data were reduced using DoPhot.

Error bars from each observatory were rescaled in order to impose
$\chi^2/{\rm dof}\simeq 1$ based on the best-fit model.

{\section{Lightcurve Analysis}
\label{sec:anal}}

The lightcurves were fitted to five-parameter models (plus two
parameters for each observatory $i$, the source flux $F_{S,i}$ and the blended
flux $F_{B,i}$),
\begin{equation}
F_i(t) = F_{S,i}A_i(t;t_0,u_0,t_\e,\pi_{\e,N},\pi_{\e,E}) + F_{B,i}
\label{eqn:foft}
\end{equation}
where \citep{pac86}
\begin{equation}
A_i(u_i) = {u^2_i + 2\over \sqrt{u^4_i + 4 u^2_i}};
\quad
u^2_i \equiv (\tau^\prime_i)^2 + (\beta^\prime_i)^2;
\quad
\tau^\prime = {t-t_0\over t_\e} + \Delta\tau_i(t);
\quad
\beta^\prime = u_0 + \Delta\beta_i(t)
\label{eqn:pac}
\end{equation}
and where, following closely the procedure 
based upon a geocentric point of view outlined in \cite{gould04},
$(\Delta\tau_i(t),\Delta\beta_i(t))$ is the apparent lens-source
offset in the Einstein ring relative to a uniform trajectory,
as seen by the $i$th observatory, due to the physical offset
(in AU) of this observatory from a rectilinear trajectory defined
by Earth's position and velocity vectors 
at the peak of the event, ($t_{0,\oplus}$).

The physical offset of the observatory
$\Delta {\bf p}_i(t) = (\Delta p_{i,N}(t),\Delta p_{i,E}(t))$
is the sum of two terms
\begin{equation}
{\bf p}_i(t) = {\bf s}(t) + {\bf t}_i(t)\,.
\label{eq:delta_p}
\end{equation}
The first term (common to all observatories) is the offset of the
apparent position of the Sun (projected on the plane of the sky)
relative to where it would be if Earth were in rectilinear motion
(see \citealt{gould04}, and specifically his Figure~(2)).  
The second term (called ``$\bf t$'' because it usually reflects
so-called ``terrestrial parallax'', as opposed to the ``orbital parallax'') 
is the projected separation of
Earth's center from the $i$th observatory.  Both terms are, by
convention, scaled to 1 AU.  The sign convention is due to the
explicitly ``geocentric'' framework.  For terrestrial observatories,
for which we use Earth's ephemerides and the location 
of each observatory relative to Earth's center,
$|{\bf t}_i|\ll 1$, although this term can in principle be important, 
particularly for high
magnification events \citep{gould97,ob07224,ob08279,gouldyee13}.
For {\it Spitzer} $|{\bf t}_i|\sim {\cal O}(1)$,
with the spacecraft position relative to Earth being 
available as a function of time from the Horizons Ephemeris 
System\footnote{http://ssd.jpl.nasa.gov/?horizons.}.
Then, in analogy to Equation (8) of \cite{gould04}
\begin{equation}
\Delta\tau_i = -{\Delta p_{i,N}\pi_{\e,N} + \Delta p_{i,E}\pi_{\e,E}\over\au}
\qquad
\Delta\beta_i = -{-\Delta p_{i,E}\pi_{\e,N} + \Delta p_{i,N}\pi_{\e,E}\over\au}\,.
\label{eqn:dtaudbeta}
\end{equation}

As discussed in Section~\ref{sec:color}, for each event 
(except OGLE-2014-BLG-0337) we measured the instrumental source color
in either $(V-I)_S$ or $(I-H)_S$.  We then determined the $(I-[3.6])_S$
color using a $VI[3.6]$ or $IH[3.6]$ color-color relation derived from
field stars.  These estimates typically have errors of 
$\sigma_{I-[3.6]}=$0.06--0.1 mag,
although they are larger in a few cases.  These color measurements
were then incorporated into the fit by
\begin{equation}
\chi^2_{\rm color} = {[(I-[3.6])-2.5\log(F_{S,Spitzer}/F_{S,\rm OGLE})]^2\over
\sigma^2_{I-[3.6]}}.
\label{eqn:colterm}
\end{equation}
In most cases inclusion of this term made almost no difference, generally
because the fit values of $F_{S,Spitzer}$ and $F_{S,\rm OGLE}$ already
yielded an $(I-[3.6])_S$ color  that was consistent with the one
derived from the color-color diagram.  However, in a few cases, particularly
when the {\it Spitzer} observations covered only a fragment of the
\citet{pac86} curve, this constraint proved to be important.

To locate the four solutions (with different parallax vectors $\bpi_\e$)
that are predicted from theory \citep{refsdal66,gould94},
we begin by fitting the ground-based lightcurve
to the standard \citet{pac86} three parameters $(t_{0,\oplus},u_{0,\oplus}, t_\e)$,
i.e., without parallax.  We then add in {\it Spitzer} data and include
two additional parameters $\bpi_\e$ and apply Newton's method 
\citep{simpson}.  This quickly locates the $\Delta u_{0,-,+}$ solution.
We then reverse the signs of $(u_0,\pi_{\e,N})$ \citep{ob09020} and
again apply Newton's method, which locates the $\Delta u_{0,-,-}$ solution.
We then take the original solution, put in a large value for $\pi_{\e,N}$,
and apply Newton's method,
which locates the $\Delta u_{0,+,+}$ solution, and finally we reverse the
signs of $(u_0,\pi_{\e,N})$ for this solution and again apply Newton's method 
to obtain the fourth solution.

The only event for which this procedure failed was OGLE-2014-BLG-1049.
The reason for the failure is that the event was high magnification 
as seen from Earth ($u_{0,\oplus}<0.01$) and was also high magnification
as seen from {\it Spitzer}.  However, because the first {\it Spitzer}
point was one day {\it after} peak, $u_{0,{\rm spitzer}}$ is consistent
with both zero and values that are significantly larger than $u_{0,\oplus}$.
These characteristics lead to a merger of the two solutions $\Delta u_{0,\pm,+}$
and also a merger of the solutions $\Delta u_{0,\pm,-}$.  Nevertheless, although
the merged solutions are unstable to Newton's method, they have quite well
behaved minima and constitute an interesting limiting case of the standard
four-fold degeneracy.

Table~\ref{tab:evt} lists the fitted parameters for each of the four solutions
for each of the 21 events.
The $\Delta\chi^2$ offset between each of the other three solutions and
the best one is shown in the second column.

An additional analysis we might in principle
address is related to the determination
of the parallax from ground-based data alone.
While formally it is extremely straightforward to fit
the lightcurves after excluding the {\it Spitzer} data
(and indeed, within our fit procedure, the effect of parallax
for ground-based data from orbital motion is automatically included), 
historical experience with ground-based parallax measurements shows that a
more cautious approach is required.  In contrast to space-based
parallaxes, in which the signal derives from obvious differences
in the peaks of the event as seen from well-separated observatories,
ground-based parallaxes derive from subtle distortions of the
lightcurves.  These can be caused or corrupted by ``xallarap''
(binary motion of the source during the event), very small distortions
due to unrecognized binary lenses, or just systematics in the
data.  These problems can be mitigated by the presence of well-understood
structures in the lightcurve for events that contain a planet 
(e.g., \citealt{muraki11}), but for point-lens events, which are otherwise
featureless, ground-based parallaxes are especially prone to such
corruption. Indeed, in the only 
systematic study of point-lens ground-based parallaxes
\citep{poindexter05}, even within a restricted
sample of parallax detections with $\Delta\chi^2>100$,
there was a strong evidence for xallarap in 23\% of cases.  
As described in some detail by \cite{poindexter05}, the
tests for xallarap (and related systematics) are quite involved
and are well beyond the scope of the present work, which relies
on much more straightforward space-based parallaxes.

{\section{Visual Representations of Solutions}
\label{sec:visual}}

Figure~\ref{fig:lc} gives a visual representation of all the key information for
20 of the 21 events  (except OGLE-2014-BLG-1049).
For each event, the upper panel shows the lightcurve
data from all observatories.  All have been aligned to OGLE fluxes
(and then converted to OGLE magnitudes) in the standard fashion.  That is,
\begin{equation}
F_{\rm OGLE,sys} = (F - F_{B,i}){F_{S,OGLE}\over F_{S,i}} + F_{B,\rm OGLE}
\label{eqn:transform}
\end{equation}
where the $F_S$ and $F_B$ are determined from the fit.  This panel also shows
the model(s), i.e., the model lightcurve as seen from Earth and from 
{\it Spitzer}.  Note that the model is extended beyond the range of
{\it Spitzer} observations although {\it Spitzer} could not actually observe
the events at these times due to Sun-angle restrictions.  The $\Delta\chi^2$
values for the four solutions are listed above this panel, always in the
same order $(-+,--,++,+-)$.  The next panel shows the residuals.

The lower two panels show two different representations of the four
parallax solutions.  In each case, the solutions are color coded in
order of increasing $\chi^2$, namely black, red, cyan and blue.  The right
panel shows the $\bpi_{\e,\geo}$ vectors and error ellipses
in the geocentric frame, i.e., those that are directly returned by the
fit.  As described below, the $\bpi_{\e,\hel}$ vectors
would have exactly
the same lengths but slightly different directions compared to the 
$\bpi_{\e,\geo}$ vectors that are shown.

In the left hand panel, we show the heliocentric projected velocities
$\tilde\bv_\hel$, defined as
\begin{equation}
\tilde\bv_\hel = \tilde\bv_\geo + \bv_{\oplus,\perp};
\qquad
\tilde\bv_\geo = {\bpi_{\e,\geo}\over\pi_\e^2}\,{\au\over t_{\e,\geo}},
\label{eqn:vhel}
\end{equation}
where $\bv_{\oplus,\perp}$ is the velocity of Earth projected on the
plane of the sky and evaluated at $t_{0,\oplus}$.  While this quantity
varies slightly from event to event in the sample, most are quite
close to $\bv_{\oplus,\perp}{\rm (N,E)}\sim(0,30)\,\kms$.  Hence,
$\tilde\bv_\geo$ can easily be estimated from these diagrams by eye
simply by displacing all vectors by $30\,\kms$ to the West.
(For completeness
we note that $\bpi_{\e,\hel} = (\tilde \bv_\hel/\tilde v_\hel)\pi_{\e,\geo}$.)

These diagrams can be used to judge the relative plausibility of
the four solutions.  Consider, for example, OGLE-2014-BLG-0678.  
The {\it Spitzer} and ground-based lightcurves are very similar,
i.e., similar $t_0$ and $u_0$.  This is what would be expected if 
$\pi_\e$ were very small, and indeed such small $\pi_\e$ is apparent
for the black $(-+)$ and red $(--)$ solutions in the lower-right panel.
However, this panel also shows the two $(+\pm)$ solutions, which have
similarly small $\Delta t_0$ (so similar $\pi_{\e,E}$) to the $(-\pm)$ solutions,
but very different $\Delta u_0$ (so $\pi_{\e,N}$).  These correspond roughly
to $\Delta u_0 \sim \pm 2 u_0$.  One of these solutions can clearly be
ruled out by its high $\Delta\chi^2(++) = 24.6$, but the other is only
slightly disfavored, $\Delta\chi^2(+-) = 3.7$.

Nevertheless, following the previously noted argument of James Rich 
(Section~\ref{sec:rich}) both of the $(+\pm)$ solutions 
for OGLE-2014-BLG-0678 are highly disfavored.
To make the general argument more concrete, we present a ``worked
example'' for this case.

We first note the values, $|\Delta \tau|\sim 0.04$, 
$|\Delta u_{0,-,\pm}|\sim 0.07$, and $u_{0,\oplus}\sim 0.43$
(here $\Delta\tau\equiv \Delta t_0/t_{\e,\oplus}$),
with therefore $|\Delta \tau| \sim |\Delta u_{0,-,\pm}|$
and additionally $|\Delta u_{0,-,\pm}| \ll |u_{0,\oplus}|$.
As expected on general grounds we also find
$|\Delta u_{0,+,\pm}|\sim 0.8 \sim 2 |u_{0,\oplus}|$.
We therefore fall within the situation
for which the Rich argument applies, $\pi_{\e,-}\ll\pi_{\e,+}$,
and we can conclude that the $\pi_{\e,-}$ solution is correct.
Indeed, according to Equations~(\ref{eq:rich_eps})
and (\ref{eq:rich_prob}), if $\pi_{\e,+}$ were correct,
with $\epsilon_0=0.1$ in this case, the probability
of finding such a small ratio would be about $P\sim 0.5\%$.

More generally, we evaulate the impact of the Rich argument
using Equation~(\ref{eq:rich_prob}) which was derived in Section~\ref{sec:rich}.
The argument applies strongly (in the sense that $\pi_{\e,+\pm}\ga 8\pi_{\e,-,\pm}$,
i.e., $P<1/128$) to a total of 10 events.
Of these 10, the argument is strongly confirmed by $\Delta\chi^2>16$ for
three cases
(OGLE-2014-BLG-0419,
OGLE-2014-BLG-0641,
OGLE-2014-BLG-0667), and
moderately ($\Delta\chi^2>9$) and
marginally ($\Delta\chi^2>4$) confirmed for one each,
OGLE-2014-BLG-0752 and OGLE-2014-BLG-0670, respectively.
For four other cases 
OGLE-2014-BLG-0678, 
OGLE-2014-BLG-0866, 
OGLE-2014-BLG-0979, OGLE-2014-BLG-1147)
there is no significant information from $\Delta\chi^2$.  Finally,
there is one case (OGLE-2014-BLG-0772)
for which Rich's argument is marginally contradicted by $\Delta\chi^2=7.2$.

The argument applies with moderate strength 
($2.5\la \pi_{\e,+\pm}/\pi_{\e,-,\pm}\la 8$)
to five events
(OGLE-2014-BLG-0337, OGLE-2014-BLG-0494, OGLE-2014-BLG-0805,
OGLE-2014-BLG-0807, OGLE-2014-BLG-0944).  
There is strong confirmation from $\Delta\chi^2$
for the third of these, strong contradiction for the second,  and no 
information from the remaining three.

We conclude that this argument can be reliably applied only to strong
cases, and should be applied to moderate cases only when significantly
confirmed by $\Delta\chi^2$.  In particular we note that of the six
cases for which the Rich argument was strongly tested by $\Delta\chi^2$, the
only case for which it was contradicted was OGLE-2014-BLG-0337, i.e.,
a moderate case with $\pi_{\e,+}/\pi_{\e,-}\sim 2.5$.

Finally, we note that we have included OGLE-2014-BLG-0939 in 
Figure~\ref{fig:lc}, which was
previously analyzed by \citet{ob140939}, to allow easy comparison with 
the other isolated-lens events.  At the level of analysis of the current paper
this would be ranked as a case for which the Rich argument is moderately
applicable and is marginally confirmed by $\chi^2$. 
In fact, the source proper-motion
measurement carried out by \citet{ob140939}
actually strongly confirms the $(--)$ solution.

Figure~\ref{fig:1049} illustrates the special case of OGLE-2014-BLG-1049.
The Earth-based lightcurve (upper panel) is quite well determined by
the combination of OGLE and PLANET SAAO data, which latter begin just 7 hours
after the high-magnification ($u_{0,\oplus}=0.01$, $A_{{\rm max},\oplus}=100$)
peak.  By contrast, the {\it Spitzer} data, which begin about 13 hours
later, leave the peak magnification as seen by {\it Spitzer} relatively
unconstrained.  In particular, $u_{0,\rm Spitzer}$ is consistent
with zero, implying that there are a continuum of viable solutions
across this ``boundary'' from $u_{0,\rm Spitzer}>0$ to 
$u_{0,\rm Spitzer}<0$. and hence a merger of the $\Delta u_{0,\pm,+}$ solutions
(also of the $\Delta u_{0,\pm,-}$ solutions).  The $\bpi_{\e,\geo}$ distribution
for the $\Delta u_{0,\pm,+}$ solutions is shown in the lower right panel
and the corresponding $\tilde\bv_\hel$ distribution at the lower left.
The $\Delta u_{0,\pm,-}$ solutions (not shown) look extremely similar
and have a nearly identical $\chi^2$ minimum.

{\section{Distribution of Lens Distances}
\label{sec:distribution}}

For each of the 22 isolated-lens events (21 analyzed here plus 
OGLE-2014-BLG-0939), we calculate the relative likelihood of the
lens being at different distances and display our results in
Figure~\ref{fig:cum}.  As explained below, the abscissa is not the lens distance
but rather
\begin{equation}
D\equiv {\kpc\over \pi_\rel/{\rm mas} + 1/8.3}
\label{eqn:ddef}
\end{equation}
which has limiting forms
\begin{equation}
D\rightarrow D_L \quad (D_L\la D_S/2);
\qquad
(8.3\,\kpc - D)\rightarrow (D_S -D_L) \quad (D_L\ga D_S/2).
\label{eqn:dlim}
\end{equation}

The probability distribution is calculated using a restricted set
of Bayesian priors, i.e., primarily {\it kinematic} priors, combined
with the measured values of $\tilde \bv_\hel$ and $\pi_\e$ as well
as discrimination among the four solutions based on $\chi^2$ and the
Rich argument.  That is, there are essentially three factors
(in addition to the lightcurve-based measurements):
phase space density, $\Delta\chi^2$ (displayed above the lightcurve
for each event in Figure~\ref{fig:lc} and color-coded in the bottom
panels), and the Rich argument.
As discussed in Section~\ref{sec:visual}, the last was applied 
by suppressing the $(+\pm)$ solutions, but only
for the 10 ``strong cases'' listed there.

The phase-space density combines the observed value of $\tilde\bv_\hel$ with
the kinematic priors.  It is computed as an integral along the line
of sight, with four factors derived from the generic 
rate equation``$\Gamma=n\sigma v$''.
The first is a volume element $D_L^2\Delta D_L$.  The second is the 
value of the expected $\tilde \bv$ distribution at the measured value,
which we describe below.  The third is the ``cross section''
which is $2\theta_\e = 2\pi_\rel/\pi_\e$.  Since $\pi_\e$ is constant
along the integral, this factor is effectively $\propto\pi_\rel$.  The
fourth is the ``velocity'' $\mu = \pi_\rel\tilde v/\au$.  Again, since
$\tilde v$ is constant, this term is also $\propto \pi_\rel$.  Hence,
ignoring for the moment the projected-velocity distribution term,
the integrand is just 
$(\pi_\rel D_L)^2 \rightarrow (1 - D_L/D_S)^2$,
which falls off fairly slowly in the disk and then drops rapidly in the bulge.

All sources were assumed to be in the bulge, and to have an isotropic
proper-motion dispersion in the bulge frame of $\sigma_\mu = 3.0\,\masyr$
(corresponding to $\sim 120\,\kms$)
in each direction.
Bulge lenses were assumed to have the same proper-motion distribution.
Disk lenses were assumed to be moving with peculiar motions of dispersions
$18\,\kms$ and $33\,\kms$ in the vertical and rotation directions relative
to a flat rotation curve at $v_{\rm rot}=240\,\kms$.  The Sun was taken
to be moving at $7\,\kms$ and $12\,\kms$ relative to the same rotation
curve.  

For disk lenses we simply assumed that the source was at $D_S=8.3\,\kpc$.
Of course, these sources are actually at a range of distances, and the
mean distance varies as a function of Galactic longitude due to the
tilt of the Galactic bar.  
However, to first order, our determinations are sensitive
only to $\pi_\rel$ (rather than to $D_L$ and $D_S$ separately), so stepping
over a discrete set of $D_S$ would just yield extremely similar
distributions in $\pi_\rel$.  It is for this reason that we report
the quantity ``$D$'' in Figure~(\ref{fig:cum}), which is a monotonic function of
the quantity ($\pi_\rel$) 
that we are actually measuring (Equation~(\ref{eqn:ddef})).  The
reasons for reporting $D$ rather than $\pi_\rel$ itself are two-fold.
First, $\pi_\rel$ is not commonly used as an independent variable, and
hence intuition about it is not widespread.  This is particularly
problematic because many lenses
would be bunched up at low $\pi_\rel$.  More importantly however,
the figure as plotted gives direct information about $D_L$ for
essentially all lenses in the disk (just from the value of $D$),
and it gives direct information about the distance from the lens
to the source for all lenses in or near the bulge from
$D_S - D_L\simeq 8.3\,\kpc - D$.

For bulge lenses we conducted an integral over lens distances for each
value of ``$D$'' by first translating this quantity into $\pi_\rel$ and
then holding this fixed while allowing $D_L$ to vary.  We adopted
an $r^{-2}$ profile for the bulge, flattened in the vertical direction 
by a factor 0.6, and we truncated it at $2\,\kpc$.  That is, in the
above integrals, we weighted by the product of the densities of the
lenses and sources, according to the Galactic coordinates of the source.

Since $\pi_\e$ is measured, each $\pi_\rel$ implies a mass 
$M=\pi_\rel/\kappa\pi_\e^2$.  We truncated the bulge lenses at
$M>1.1\,M_\odot$ and the disk lenses at $M>1.5\,M_\odot$ due to the
paucity of such stars in each population.
There may be additional
modest constraints on lens mass from (lack of) blended light but
we did not attempt to evaluate these.

As can be seen from Figure~\ref{fig:cum}, the great majority of disk lenses have
distance distributions that are relatively compact and characterized
by a single peak.  This can be understood by inspection of Figure~\ref{fig:lc}.
In many cases, one solution is strongly preferred by $\chi^2$.
OGLE-2014-BLG-0678 provides a good example for which there is no
strong preference in $\chi^2$ between the two allowed solutions.  
However, one of the two solutions (black)
is closely aligned with the direction of Galactic rotation (roughly
$30^\circ$ East of North) and so is strongly favored by the kinematic
priors.  The second (red) solution then contributes 
almost nothing to the total probability.  OGLE-2014-BLG-0670
provides another instructive case.  Here the kinematically preferred
solution (red) is marginally disfavored by $\Delta\chi^2=3.2$.  These
two factors roughly cancel, but the two solutions predict very
similar distance distributions, so the combined probability
distribution function is only slightly broadened by the ambiguity.
In fact, of all the lenses in the sample, 
there are only two that are double-peaked:
OGLE-2014-BLG-0944 and OGLE-2014-BLG-1021.  In both case, the $(-\pm)$
solutions correspond to bulge lenses while the $(+\pm)$ solutions
correspond to disk lenses.  And in both cases, $\chi^2$ does very
little to discriminate between possible solutions.  Hence we treat
the bulge and disk solutions as equally likely in each case.  The
resulting double-peaked probability distributions are shown
as bold-dashed curves in Figure~\ref{fig:cum}.
Of the six other bulge-lens events,
one has somewhat double-peaked features due to slightly different
$\tilde v$ and the fact that the direction of $\tilde \bv$ 
does not differentiate between solutions for bulge lenses (because of the
assumed isotropy of proper motions).

Note that the Galactic model used for the distance measurements is 
simplified in a number of respects.  First, there is no weighting
by an assumed lens mass function, which is equivalent to assuming
a flat prior in log mass.  Second, for the disk lenses, there is no
weighting by stellar density, which is equivalent to assuming that
the declining density with distance from the plane due to the vertical
scale height exactly cancels the increasing density as one approaches
the Galactic center due to the radial scale length.  And of course,
we do not attempt to model even finer details, such as varying 
velocity dispersion, changing scale heights etc.

We do not develop more sophisticated models
for three reasons.  First, we wish to demonstrate the power of kinematic
priors (combined with $\tilde\bv$ measurements) alone to constrain the 
distances to individual lenses.  This point has been made before 
theoretically \citep{han95}, but has never been demonstrated practically.

Second, the distance measurements individually, and especially cumulatively,
are robust against modest changes in assumptions.  
For example, if $v_{\rm rot}$ is changed from
$240\,\kms$ to $220\,\kms$, then the resulting version of Figure~\ref{fig:cum}
is indistinguishable by eye from the current one.  As another example,
we have recomputed the distance distributions in Figure~\ref{fig:cum}
using more realistic mass priors $d N/d\ln M\propto M^{-x}$, with
$x=0.3$ (1.3) for $M> (<) 0.5\,M_\odot$ for the disk and
$x=1$ (0.3) for $M> (<) 0.7\,M_\odot$ for the bulge.  We plot 
the resulting cumulative distributions in Figure~\ref{fig:cum} with (solid)
and without (bold) mass-function priors 
and note
that they hardly differ.
The reason for this is that over the regions of parameter space permitted
by the kinematic priors, the mass priors generally do not vary very much.

Third, the proper
context to study the impact of model variations is within a determination
of the Galactic distribution of planets.  As we discuss in 
Section~\ref{sec:path} immediately below, such a measurement will require
additional data.

{\section{Pathway to Galactic Distribution of Planets}
\label{sec:path}}

Figure~\ref{fig:cum} shows the cumulative distribution of $D$ (monotonic 
function of $\pi_\rel$), constructed by adding together all the lens probability
distributions and normalizing to unity. The
position of the one planet in the {\it Spitzer} sample
(OGLE-2014-BLG-0124, \citealt{ob140124}) is also shown.  Of course, nothing
can be said about the Galactic distribution of planets based on a single
planet.  However, as emphasized in Section~\ref{sec:intro}, events can
be added from future observing campaigns by either {\it Spitzer} or
other space observatories, with the isolated lenses forming the cumulative
distribution function and the planetary events being used to measure
the distance distribution of planets relative to this cumulative distribution.

Note that, in general, the individual distance measurements for the
planetary events will be more accurate than for the isolated-lens events.
This is because the former will mostly have measurements of $\theta_\e$
(from caustic crossings and/or approaches) and
thus $\pi_\rel = \theta_\e\pi_\e$, while the latter will have distance
estimates based on measured $\tilde\bv_\hel$ combined with kinematic
priors.  However, because there are many more isolated-lens events
than planetary events and because the kinematic distance estimates
for the isolated lenses are relatively accurate (see Figure~\ref{fig:cum}), 
uncertainties
in the cumulative distribution function will not contribute much to
uncertainties in the overall measurement.  Rather, the precision of
measurement of the Galactic distribution of planets will depend directly
on how many planets are detected in space-based parallax surveys.

There are essentially two ways to increase the number of planets detected
in space-based campaigns.  The first is simply to observe in additional
years and/or with additional satellites.  Both {\it Spitzer} and {\it Kepler}
(in its K2 mode) are quite well suited to this task.  The second is to
make more intensive use of the time available for Galactic bulge observations.
In the case of {\it Kepler} this is an automatic feature since {\it Kepler}
observes its targets almost continuously as a matter of course.  For
{\it Spitzer} more intensive observing can increase the number of
planetary detections in two ways:  first by allowing more events to
be monitored and second by detecting planets from space that are not
detected from the ground.  Because the spacecraft probes a region
of the Einstein ring that is more or less separated from the one
seen from the ground, it can observe planetary caustics that are not
seen from the ground \citep{gouldhorne}.  However, this requires
that the events be observed several to many times per day as
compared to roughly once per day in the present campaign.

We note that roughly 30\% of the lenses in our sample are in the bulge
compared to roughly 60\% expected for an unbiased sample of lensing events.
Qualitatively, the reason for this is clear: the delay between recognition
of the events and uploading coordinates to the spacecraft biases the
sample to long events, which are preferentially in the disk.  The
same bias (for somewhat different reasons) affects the \citet{gould10}
sample of high-magnification events.  

This bias in the sample of underlying events does not in any way bias
the measurement of the Galactic distribution of planets
because the planetary events are subject to the same
selection effects.  However, to
the extent that bulge events are underrepresented in the sample, it 
does mean that more planetary detections will be needed to measure the
bulge-versus-disk fractions compared to what would be the case if there
were more bulge events.  Thus, it is important to develop more aggressive
methods of identifying shorter events in time to upload coordinates,
to the extent that this is possible.

Finally, we note the Galactic distribution of planets must be determined
from the cumulative distribution (with distance)  of {\it planet sensitivity}
of events with parallaxes, not simply the cumulative distribution of
the events themselves (as in Figure~\ref{fig:cum}).  
Such planet-sensitivity calculations
are an essential feature of all microlens planet frequency analyses.
See, e.g., Figure 8 of \citet{gaudi02} or Figures 2--4 of \citet{gould10}.
Since microlens planet sensitivity is a function of both planet-star mass ratio
$q$ and normalized separation $s$, such studies of the Galactic distribution
of planets
can in principle also yield functions of these variables.  At the first
stages, however, all that will be accessible is the distribution of a planet
frequency that is suitably averaged over $q$ and $s$.

{\section{Conclusions
\label{sec:conclude}}

We have measured the microlens parallaxes of 21 events that were discovered
by OGLE and observed by {\it Spitzer}, which was located $\sim 1\,\au$
West of Earth in projection.  We used kinematic priors based on a Galactic
model to estimate distances to each of the lenses.  In the great majority
of cases, these distributions are well localized, as illustrated in Figure~\ref{fig:cum}.
Such localization was not guaranteed in advance because the lens distances
are subject to a well-known four-fold degeneracy \citep{refsdal66,gould94}.

In the case of 10 of the 21 events, we were able to break the key
element of this degeneracy by quantifying and testing an argument
originally given by James Rich (circa 1997, private communication).
In its quantified form, this states that,
provided that $\pi_{\e,-}\ll \pi_{\e,+}$,
the $\Delta u_{0,+,\pm}$ solutions
(in which the source appears on the opposite side of the lens as
seen from Earth and from the satellite) are less probable than
the $\Delta u_{0,-,\pm}$ solutions by a factor $\sim (\pi_{\e,-}/\pi_{\e,+})^2/2$.
The remaining degeneracy within the $\Delta u_{0,-,\pm}$ solutions is relatively
unimportant because it leads to similar distance estimates and because,
at least for disk lenses, the kinematic priors usually discriminate between 
these two solutions.  As demonstrated by Figure~\ref{fig:cum}, only two 
of the 21
events have substantially extended probability distributions of the distance
variable $D\equiv \kpc/(\pi_\rel/{\rm mas} + 1/8.3)$.

We have shown that an accurate cumulative distribution function of lens
distances can be constructed from our sample.  This means that the distances
of planets detected from the same program can be used to determine the
Galactic distribution of planets.  That is, the {\it Spitzer} sample
is a fair parent sample for the planets detected, even though the sample
itself is biased toward longer events (and so disk lenses).  The reason
that this sample is nevertheless fair is that planetary events and
the non-planetary events suffer exactly the same bias because the
planetary nature of the events is not known at the time the decision
is made to observe them (e.g., \citealt{ob140124}).
This means that this sample can be combined
with future samples, including those observed in future years by 
{\it Spitzer} and {\it Kepler}, even though the selection biases of these
samples are likely to differ radically.

\acknowledgments

Work by CAB  was carried out in part at the Jet Propulsion Laboratory (JPL),
California Institute of Technology, under a contract with the National
Aeronautics and Space Administration.
Work by JCY, AG, and SC was supported by JPL grant 1500811.
AG and BSG were supported by NSF grant AST 1103471.
AG, BSG, and RWP were supported by NASA grant NNX12AB99G.
The OGLE project has received funding from the European Research Council
under the European Community's Seventh Framework Programme
(FP7/2007-2013) / ERC grant agreement no. 246678 to AU.
Work by JCY was
performed under contract with the California Institute of Technology
(Caltech)/Jet Propulsion Laboratory (JPL) funded by NASA through the
Sagan Fellowship Program executed by the NASA Exoplanet Science
Institute.
Work by DDP and KL were supported by the University of Rijeka 
project 13.12.1.3.02. 
Work by TS is supported by grants JSPS23103002, JSPS24253004, and JSPS26247023.
Work by IAB was supported by the Marsden Fund of the Royal Society of New
Zealand, contract no. MAU1104.
The MOA project is supported by the grant JSPS25103508 and 23340064.
Work by DM is supported by the I-CORE program of the Planning and
Budgeting Committee and the Israel Science Foundation, Grant 1829/12. DM
and AG acknowledge support by the US-Israel Binational Science Foundation.
The operation of the Danish 1.54m telescope at ESO’s La Silla Observatory is
financed by a grant to UGJ from the Danish Natural Science Foundation (FNU).
This publication was made possible by NPRP grants 09 - 476 - 1 - 078 and
X-019-1-006 from the Qatar National Research Fund (a member of Qatar
Foundation). OW (FNRS research fellow) and J Surdej acknowledge support from
the Communaut\'e fran\c caise de Belgique - Actions de recherche
concert\'ees - Acad\'emie Wallonie-Europe. MH acknowledges support from the
Villum Foundation. CS received funding from the European Union Seventh
Framework Programme (FP7/2007-2013) under grant agreement no. 268421.
This work is based in part on observations made with the Spitzer Space
Telescope, which is operated by the Jet Propulsion Laboratory,
California Institute of Technology under a contract with NASA.
The research described in this publication was carried out in part at the
Jet Propulsion Laboratory, California Institute of Technology, under a
contract with the National Aeronautics and Space Administration.

\clearpage
\begin{deluxetable}{ccccrcl}
\tabletypesize{\scriptsize}
\rotate    
\tablecaption{Event parameters\label{tab:evt_param}}
\tablewidth{0pt}
\tablehead{
\colhead{Event} & \colhead{RA (J2000)} &\colhead{DEC (J2000)} &
\colhead{$\beta_\mathrm{ec}$ (J2000)}& $I-[3.6]$ & \emph{Spitzer}&ground-based data\\
\colhead{OGLE-2014-BLG-} & degree &degree &
degree&& epochs & \\
}
\startdata
 0099 &  269.607333 & -28.279833 &  -4.94030  & $-0.69\pm 0.06$&32&OGLE, MOA, Wise, RoboNet\tablenotemark{a,b,c}\,\,\,, MiNDSTEp\tablenotemark{d}\\
 0115 &  269.156917 & -28.515750 &  -5.17792  & $-0.86\pm 0.06$&22&OGLE, MOA, Wise, MiNDSTEp\tablenotemark{d}\\
 0337 &  267.841125 & -29.733250 &  -6.40733  &                &37&OGLE, Wise\\
 0419 &  269.629708 & -30.100639 &  -6.76105  & $-1.01\pm 0.07$&37&OGLE, Wise\\
 0494 &  273.191542 & -28.227139 &  -4.91827  & $-1.41\pm 0.15$&43&OGLE, MOA, RoboNet\tablenotemark{a,b,c}\,\,\,, MiNDSTEp\tablenotemark{d,e}\\
 0589 &  268.380625 & -21.014917 &   2.31661  & $ 0.54\pm 0.08$&23&OGLE, RoboNet\tablenotemark{a,b,c}\,\,\,, MiNDSTEp\tablenotemark{e}\\
 0641 &  267.682667 & -33.905972 & -10.58165  & $-1.11\pm 0.06$&28&OGLE, MOA\\
 0667 &  272.704625 & -26.418028 &  -3.10071  & $-1.20\pm 0.07$&36&OGLE, MOA\\
 0670 &  265.542000 & -33.495472 & -10.21366  & $ 1.16\pm 0.20$&20&OGLE\\
 0678 &  267.976667 & -31.903389 &  -8.57554  & $ 0.03\pm 0.10$&33&OGLE\\
 0752 &  270.657333 & -29.594694 &  -6.25600  & $-1.13\pm 0.06$&29&OGLE\\
 0772 &  265.581875 & -23.618861 &  -0.34067  & $-0.34\pm 0.07$&26&OGLE\\
 0805 &  263.152708 & -28.163667 &  -4.96693  & $ 0.34\pm 0.17$&25&OGLE, PLANET\\
 0807 &  265.186792 & -23.863722 &  -0.59693  & $-0.86\pm 0.09$&25&OGLE, PLANET\\
 0866 &  268.025458 & -23.409194 &  -0.08156  & $-0.47\pm 0.10$&25&OGLE\\
 0874 &  270.230125 & -27.545861 &  -4.20602  & $-1.26\pm 0.08$&34&OGLE, PLANET, MOA, Wise, RoboNet\tablenotemark{a,b}, MiNDSTEp\tablenotemark{d,e}\\
 0944 &  263.204125 & -28.439028 &  -5.23984  & $ 0.42\pm 0.15$&19&OGLE, MiNDSTEp\tablenotemark{e} \\
 0979 &  267.682500 & -35.709139 & -12.38457  & $-1.50\pm 0.12$&13&OGLE, PLANET\\
 1021 &  264.315042 & -29.194722 &  -5.95271  & $ 0.02\pm 0.10$&18&OGLE, PLANET\\
 1049 &  274.107125 & -31.012333 &  -7.72275  & $-2.31\pm 0.06$&19&OGLE, PLANET\\
 1147 &  261.205875 & -29.600222 &  -6.49370  & $-0.35\pm 0.08$& 7&OGLE\\
\enddata
\tablecomments{For the ensemble of the 21 events we report the 
name, according to the OGLE naming scheme, the coordinates, the 
instrumental color, $I-[3.6]$, evaluated as discussed in the text, 
the number of epochs
of \emph{Spitzer} observations and the ground-based data used for the analysis.
The reported instrumental colors are suitable for {\it Spitzer} data
reduced by a PRFs-based analysis; for data reduced by aperture
photometry we use an aperture correction factor.}
\tablenotetext{a}{Siding Spring LCOGT telescope (Australia).}
\tablenotetext{b}{Sutherland LCOGT telescope (South Africa).}
\tablenotetext{c}{Cerro Tololo LCOGT telescope (Chile).}
\tablenotetext{d}{Danish telescope, La Silla (Chile).}
\tablenotetext{e}{Salerno University Telescope (Italy).}
\end{deluxetable}


\clearpage
\begin{deluxetable}{crrrrrrrrrrrr}
\tabletypesize{\scriptsize}
\rotate
\tablecaption{Event fit parameters\label{tab:evt}}
\tablewidth{0pt}
\tablehead{
\colhead{Event} &\colhead{$\Delta\chi^2$} &\colhead{$t_0-6800$} &\colhead{$u_0$} &
\colhead{$t_\mathrm{E}$} & \colhead{$\pi_\mathrm{E,N}$} &\colhead{$\pi_\mathrm{E,E}$} &
\colhead{${\tilde v}_\mathrm{hel, N}$} &\colhead{${\tilde v}_\mathrm{hel, E}$} &
\colhead{$I_\mathrm{OGLE}$} &\colhead{$f_\mathrm{OGLE}$\tablenotemark{a}} &
\colhead{$mag_\mathrm{Spitzer}$\tablenotemark{b}} &\colhead{$f_\mathrm{Spitzer}$\tablenotemark{a}}\\
\colhead{OGLE-2014-BLG-} &\colhead{} &\colhead{HJD-2450000} &\colhead{}&
\colhead{day} & \colhead{} &\colhead{} &
\colhead{km/s} &\colhead{km/s} &
\colhead{} &\colhead{} &
\colhead{} &\colhead{}
}
\startdata
 0099 &   17.3 & 76.910 &  0.3828 & 116.2 & -0.0823 &  0.2060 &   -26.7 &    82.9 & 16.831 &  0.147 & 17.849 &  0.230\\
          &        &  0.383 &  0.0067 &   1.3 &  0.0045 &  0.0033 &     1.0 &     1.5 &  0.026 &  0.028 &  0.078 &  0.161\\
          &    0.0 & 76.920 & -0.4075 & 111.1 &  0.1092 &  0.2157 &    27.3 &    78.0 & 16.734 &  0.049 & 17.748 &  0.140\\
          &        & -0.407 &  0.0057 &   1.0 &  0.0056 &  0.0036 &     1.0 &     1.4 &  0.021 &  0.021 &  0.071 &  0.137\\
          &  241.5 & 76.949 &  0.2033 & 178.2 & -0.2468 &  0.1594 &   -29.6 &    38.4 & 17.728 &  1.618 & 19.042 &  3.534\\
          &        &  0.203 &  0.0023 &   1.6 &  0.0025 &  0.0022 &     0.2 &     0.2 &  0.015 &  0.035 &  0.047 &  0.385\\
          &  203.0 & 76.713 & -0.2962 & 127.1 &  0.3698 &  0.1551 &    29.5 &    33.6 & 17.214 &  0.632 & 18.558 &  1.368\\
          &        & -0.296 &  0.0029 &   0.9 &  0.0039 &  0.0024 &     0.2 &     0.2 &  0.013 &  0.020 &  0.051 &  0.254\\
 0115 &   10.1 & 59.612 &  0.2687 & 105.6 & -0.0777 &  0.1215 &   -62.6 &   121.3 & 17.303 &  0.187 & 18.170 &  0.378\\
          &        &  0.269 &  0.0036 &   1.0 &  0.0050 &  0.0039 &     2.6 &     4.7 &  0.018 &  0.020 &  0.069 &  0.196\\
          &    0.0 & 59.645 & -0.2867 & 100.5 &  0.0963 &  0.1128 &    73.9 &   113.9 & 17.212 &  0.091 & 18.094 &  0.332\\
          &        & -0.287 &  0.0033 &   0.8 &  0.0057 &  0.0040 &     3.1 &     4.6 &  0.016 &  0.016 &  0.063 &  0.179\\
          &  269.6 & 59.257 &  0.2011 & 129.0 & -0.2519 &  0.1232 &   -44.4 &    46.7 & 17.693 &  0.701 & 18.818 &  1.704\\
          &        &  0.201 &  0.0022 &   1.1 &  0.0034 &  0.0024 &     0.4 &     0.5 &  0.014 &  0.021 &  0.052 &  0.334\\
          &   94.3 & 59.395 & -0.2622 & 102.9 &  0.3416 &  0.0824 &    45.1 &    36.8 & 17.337 &  0.225 & 18.375 &  0.362\\
          &        & -0.262 &  0.0027 &   0.8 &  0.0049 &  0.0033 &     0.4 &     0.5 &  0.014 &  0.015 &  0.056 &  0.227\\
 0337 &   38.0 & 22.967 &  0.5315 &  45.0 & -0.0790 &  0.1298 &  -131.9 &   245.7 & 16.759 & -0.021 & 15.711 & -0.381\\
          &        &  0.531 &  0.0077 &   0.4 &  0.0152 &  0.0088 &    17.2 &    24.5 &  0.024 &  0.022 &  0.084 &  0.091\\
          &   19.5 & 23.016 & -0.5465 &  44.4 &  0.3282 &  0.1015 &   108.3 &    62.8 & 16.711 & -0.063 & 16.670 &  1.242\\
          &        & -0.547 &  0.0079 &   0.4 &  0.0294 &  0.0102 &     6.4 &     7.6 &  0.024 &  0.021 &  0.157 &  0.439\\
          & 1611.6 & 22.472 &  0.4691 &  48.8 & -0.6151 &  0.0715 &   -57.1 &    35.8 & 16.959 &  0.178 & 18.399 & 14.687\\
          &        &  0.469 &  0.0056 &   0.4 &  0.0091 &  0.0039 &     0.5 &     0.4 &  0.019 &  0.021 &  0.078 &  1.295\\
          &    0.0 & 23.022 & -0.6558 &  38.9 &  0.9844 & -0.2310 &    42.7 &    19.1 & 16.385 & -0.306 & 16.616 &  1.036\\
          &        & -0.656 &  0.0121 &   0.5 &  0.0295 &  0.0208 &     1.1 &     0.4 &  0.034 &  0.022 &  0.161 &  0.427\\
 0419 &    0.6 & 22.886 &  0.2401 &  48.6 & -0.0253 & -0.0334 &  -513.3 &  -648.8 & 18.202 &  0.328 & 19.391 &  0.430\\
          &        &  0.240 &  0.0048 &   0.7 &  0.0071 &  0.0033 &    86.6 &   132.2 &  0.026 &  0.031 &  0.058 &  0.136\\
          &    0.0 & 22.896 & -0.2414 &  48.4 &  0.0185 & -0.0417 &   318.3 &  -687.2 & 18.195 &  0.320 & 19.400 &  0.441\\
          &        & -0.241 &  0.0049 &   0.7 &  0.0077 &  0.0032 &    96.5 &   104.5 &  0.026 &  0.031 &  0.058 &  0.136\\
          &   94.4 & 22.763 &  0.2345 &  49.4 & -0.4832 & -0.0630 &   -70.9 &    19.8 & 18.230 &  0.363 & 19.447 &  0.578\\
          &        &  0.235 &  0.0047 &   0.7 &  0.0114 &  0.0040 &     1.1 &     0.4 &  0.026 &  0.032 &  0.058 &  0.144\\
          &   23.4 & 22.954 & -0.2549 &  46.5 &  0.4700 & -0.1851 &    68.9 &     2.1 & 18.122 &  0.234 & 19.419 &  0.325\\
          &        & -0.255 &  0.0052 &   0.6 &  0.0107 &  0.0058 &     1.0 &     0.4 &  0.027 &  0.030 &  0.059 &  0.141\\
 0494 &    5.9 & 17.310 &  0.1540 &  33.2 &  0.0381 &  0.0953 &   189.5 &   499.7 & 14.493 & -0.004 & 16.138 &  0.212\\
          &        &  0.154 &  0.0005 &   0.1 &  0.0027 &  0.0029 &     5.4 &    18.8 &  0.004 &  0.004 &  0.016 &  0.021\\
          &    0.0 & 17.306 & -0.1539 &  33.3 & -0.0048 &  0.1074 &   -20.4 &   511.8 & 14.493 & -0.003 & 16.129 &  0.204\\
          &        & -0.154 &  0.0005 &   0.1 &  0.0018 &  0.0037 &     7.3 &    17.0 &  0.004 &  0.004 &  0.017 &  0.021\\
          &  160.5 & 17.278 &  0.1547 &  33.2 & -0.4448 &  0.1101 &  -109.0 &    55.6 & 14.488 & -0.008 & 16.131 &  0.216\\
          &        &  0.155 &  0.0005 &   0.1 &  0.0028 &  0.0029 &     0.8 &     0.4 &  0.004 &  0.004 &  0.016 &  0.021\\
          &  233.9 & 17.340 & -0.1552 &  33.1 &  0.4445 & -0.0276 &   118.5 &    21.0 & 14.482 & -0.014 & 16.199 &  0.268\\
          &        & -0.155 &  0.0005 &   0.1 &  0.0032 &  0.0022 &     0.7 &     0.7 &  0.004 &  0.004 &  0.016 &  0.022\\
 0589 &    1.9 &  7.622 &  0.0518 &  33.9 &  0.3727 &  0.1983 &   106.1 &    84.3 & 16.942 & -0.052 & 16.703 &  0.118\\
          &        &  0.052 &  0.0006 &   0.3 &  0.0286 &  0.0333 &     8.0 &     9.7 &  0.012 &  0.010 &  0.070 &  0.075\\
          &    1.0 &  7.624 & -0.0516 &  34.0 & -0.3700 &  0.2040 &  -106.3 &    85.7 & 16.946 & -0.049 & 16.657 &  0.071\\
          &        & -0.052 &  0.0006 &   0.3 &  0.0293 &  0.0337 &     8.3 &     9.6 &  0.012 &  0.010 &  0.071 &  0.073\\
          &    0.0 &  7.624 &  0.0517 &  34.1 & -0.5368 &  0.1728 &   -86.6 &    55.2 & 16.945 & -0.049 & 16.639 &  0.051\\
          &        &  0.052 &  0.0006 &   0.3 &  0.0303 &  0.0350 &     4.4 &     5.9 &  0.012 &  0.010 &  0.072 &  0.072\\
          &    0.7 &  7.622 & -0.0519 &  33.9 &  0.5544 &  0.1583 &    84.4 &    51.8 & 16.939 & -0.054 & 16.701 &  0.115\\
          &        & -0.052 &  0.0006 &   0.3 &  0.0277 &  0.0335 &     4.0 &     5.3 &  0.012 &  0.010 &  0.071 &  0.075\\
 0641 &    0.0 & 46.959 &  0.5398 &  38.6 & -0.0182 &  0.0333 &  -568.5 &  1062.6 & 16.876 & -0.071 & 18.015 &  0.311\\
          &        &  0.540 &  0.0162 &   0.7 &  0.0106 &  0.0100 &   218.1 &   398.3 &  0.050 &  0.043 &  0.080 &  0.110\\
          &    0.9 & 46.974 & -0.5411 &  38.6 &  0.0202 &  0.0327 &   610.3 &  1021.2 & 16.871 & -0.075 & 17.992 &  0.277\\
          &        & -0.541 &  0.0163 &   0.7 &  0.0121 &  0.0103 &   246.1 &   418.0 &  0.050 &  0.043 &  0.077 &  0.104\\
          &  318.9 & 46.465 &  0.4434 &  42.9 & -0.6919 &  0.0509 &   -60.2 &    32.1 & 17.197 &  0.251 & 18.614 &  1.451\\
          &        &  0.443 &  0.0116 &   0.7 &  0.0194 &  0.0069 &     0.8 &     0.7 &  0.041 &  0.047 &  0.064 &  0.165\\
          &   23.7 & 46.971 & -0.5893 &  35.1 &  0.8950 & -0.1728 &    51.0 &    17.5 & 16.722 & -0.193 & 17.882 &  0.028\\
          &        & -0.589 &  0.0188 &   0.7 &  0.0286 &  0.0140 &     0.8 &     0.5 &  0.055 &  0.041 &  0.082 &  0.097\\
 0667 &    0.0 & 35.115 &  0.4672 &  32.3 &  0.1128 &  0.0386 &   426.1 &   174.8 & 16.161 &  0.022 & 17.280 &  0.224\\
          &        &  0.467 &  0.0096 &   0.4 &  0.0135 &  0.0061 &    31.2 &    46.1 &  0.032 &  0.030 &  0.075 &  0.106\\
          &    8.9 & 35.093 & -0.4656 &  32.4 & -0.1021 &  0.0588 &  -392.9 &   255.8 & 16.166 &  0.027 & 17.311 &  0.275\\
          &        & -0.466 &  0.0095 &   0.4 &  0.0146 &  0.0050 &    23.9 &    54.2 &  0.032 &  0.031 &  0.075 &  0.109\\
          &  285.3 & 34.818 &  0.4739 &  31.7 & -0.9800 & -0.0669 &   -54.9 &    25.5 & 16.137 &  0.000 & 17.485 &  0.553\\
          &        &  0.474 &  0.0102 &   0.4 &  0.0234 &  0.0086 &     0.8 &     0.4 &  0.034 &  0.031 &  0.076 &  0.132\\
          &   59.8 & 34.999 & -0.5000 &  30.6 &  1.0009 & -0.2452 &    53.8 &    16.2 & 16.050 & -0.078 & 17.261 &  0.144\\
          &        & -0.500 &  0.0111 &   0.4 &  0.0231 &  0.0122 &     0.8 &     0.3 &  0.036 &  0.031 &  0.079 &  0.109\\
 0670 &    3.2 & 15.144 &  0.7697 &  25.6 &  0.0557 &  0.0928 &   321.4 &   564.3 & 16.049 &  0.084 & 14.946 &  0.539\\
          &        &  0.770 &  0.0934 &   2.0 &  0.0697 &  0.0300 &   238.6 &   360.8 &  0.239 &  0.239 &  0.296 &  0.425\\
          &    0.0 & 15.105 & -0.7478 &  26.2 & -0.1053 &  0.1485 &  -210.1 &   325.2 & 16.106 &  0.142 & 14.777 &  0.310\\
          &        & -0.748 &  0.0890 &   2.0 &  0.0809 &  0.0436 &    82.1 &   162.6 &  0.232 &  0.243 &  0.291 &  0.355\\
          &    5.2 & 14.597 &  0.7250 &  27.0 & -1.7211 & -0.2294 &   -36.5 &    23.9 & 16.162 &  0.203 & 14.978 &  0.579\\
          &        &  0.725 &  0.0881 &   2.1 &  0.2054 &  0.0596 &     2.1 &     0.7 &  0.233 &  0.258 &  0.289 &  0.428\\
          &   32.7 & 15.481 & -0.9690 &  21.9 &  1.8265 & -0.9361 &    34.4 &    11.3 & 15.560 & -0.309 & 14.635 &  0.074\\
          &        & -0.969 &  0.1709 &   2.5 &  0.2888 &  0.2070 &     2.5 &     0.6 &  0.391 &  0.248 &  0.437 &  0.453\\
 0678 &    0.5 & 22.017 &  0.4260 &  30.3 & -0.0556 &  0.0660 &  -426.8 &   536.0 & 17.467 &  0.267 & 17.381 &  0.309\\
          &        &  0.426 &  0.0212 &   1.0 &  0.0199 &  0.0105 &    76.4 &   151.7 &  0.076 &  0.088 &  0.107 &  0.156\\
          &    0.0 & 22.034 & -0.4284 &  30.2 &  0.0810 &  0.0532 &   494.6 &   353.8 & 17.459 &  0.257 & 17.389 &  0.313\\
          &        & -0.428 &  0.0214 &   1.0 &  0.0212 &  0.0117 &    71.5 &   125.6 &  0.076 &  0.088 &  0.108 &  0.156\\
          &   24.6 & 21.863 &  0.4149 &  30.8 & -0.8523 &  0.0245 &   -65.7 &    31.0 & 17.506 &  0.312 & 17.413 &  0.383\\
          &        &  0.415 &  0.0207 &   1.0 &  0.0455 &  0.0111 &     1.9 &     0.9 &  0.075 &  0.091 &  0.107 &  0.162\\
          &    3.7 & 22.111 & -0.4541 &  29.0 &  0.8474 & -0.2429 &    65.2 &    10.5 & 17.367 &  0.155 & 17.395 &  0.215\\
          &        & -0.454 &  0.0235 &   1.0 &  0.0446 &  0.0193 &     1.8 &     0.7 &  0.081 &  0.086 &  0.111 &  0.160\\
 0752 &    0.8 & 39.353 &  0.6781 &  39.7 &  0.0669 &  0.0381 &   491.7 &   309.0 & 16.682 &  0.609 & 17.816 &  0.409\\
          &        &  0.678 &  0.0500 &   1.9 &  0.0165 &  0.0106 &    56.8 &   131.1 &  0.138 &  0.204 &  0.154 &  0.209\\
          &    0.0 & 39.311 & -0.6611 &  40.5 & -0.0664 &  0.0537 &  -389.6 &   343.6 & 16.730 &  0.681 & 17.857 &  0.474\\
          &        & -0.661 &  0.0482 &   1.9 &  0.0189 &  0.0107 &    53.7 &   114.8 &  0.134 &  0.208 &  0.149 &  0.212\\
          &  137.1 & 38.384 &  0.4633 &  49.7 & -0.8725 & -0.0944 &   -39.7 &    24.8 & 17.340 &  1.949 & 18.565 &  1.878\\
          &        &  0.463 &  0.0261 &   1.9 &  0.0469 &  0.0132 &     0.8 &     0.4 &  0.090 &  0.244 &  0.107 &  0.307\\
          &   11.8 & 38.951 & -0.7303 &  36.1 &  1.1933 & -0.3988 &    35.8 &    16.9 & 16.537 &  0.408 & 17.682 &  0.079\\
          &        & -0.730 &  0.0564 &   1.9 &  0.0831 &  0.0384 &     0.9 &     0.2 &  0.149 &  0.193 &  0.163 &  0.183\\
 0772 &    7.2 & 17.428 &  0.4645 &  26.9 & -0.0308 &  0.0161 & -1643.0 &   886.4 & 17.066 &  0.727 & 17.353 &  0.417\\
          &        &  0.464 &  0.0308 &   1.1 &  0.0124 &  0.0046 &   404.1 &   528.5 &  0.104 &  0.166 &  0.089 &  0.112\\
          &    7.2 & 17.429 & -0.4644 &  26.9 &  0.0304 &  0.0168 &  1618.8 &   923.9 & 17.066 &  0.727 & 17.353 &  0.417\\
          &        & -0.464 &  0.0308 &   1.1 &  0.0125 &  0.0046 &   394.7 &   536.3 &  0.104 &  0.166 &  0.089 &  0.112\\
          &    0.0 & 17.403 &  0.4684 &  26.8 & -0.9883 & -0.2107 &   -63.3 &    15.7 & 17.051 &  0.703 & 17.403 &  0.462\\
          &        &  0.468 &  0.0308 &   1.1 &  0.0583 &  0.0150 &     1.3 &     0.3 &  0.104 &  0.162 &  0.084 &  0.111\\
          &    1.6 & 17.415 & -0.4687 &  26.8 &  0.9903 & -0.1883 &    62.1 &    17.0 & 17.050 &  0.701 & 17.412 &  0.472\\
          &        & -0.469 &  0.0308 &   1.1 &  0.0583 &  0.0136 &     1.3 &     0.2 &  0.104 &  0.162 &  0.084 &  0.111\\
 0805 &    0.1 & 39.907 &  0.1794 &  55.8 & -0.0586 &  0.0289 &  -428.4 &   238.4 & 18.593 &  0.067 & 18.583 &  2.236\\
          &        &  0.179 &  0.0092 &   2.1 &  0.0054 &  0.0023 &    24.1 &    21.9 &  0.064 &  0.062 &  0.069 &  0.290\\
          &    0.0 & 39.919 & -0.1808 &  55.4 &  0.0609 &  0.0288 &   417.1 &   226.3 & 18.582 &  0.056 & 18.588 &  2.241\\
          &        & -0.181 &  0.0093 &   2.1 &  0.0056 &  0.0023 &    23.5 &    21.0 &  0.064 &  0.062 &  0.069 &  0.291\\
          &    1.6 & 39.865 &  0.1759 &  56.4 & -0.2461 &  0.0157 &  -126.2 &    36.1 & 18.618 &  0.092 & 18.593 &  2.283\\
          &        &  0.176 &  0.0091 &   2.1 &  0.0120 &  0.0021 &     2.0 &     1.2 &  0.064 &  0.063 &  0.069 &  0.291\\
          &    0.1 & 39.918 & -0.1823 &  54.8 &  0.2497 &  0.0120 &   124.2 &    34.2 & 18.572 &  0.047 & 18.614 &  2.299\\
          &        & -0.182 &  0.0094 &   2.1 &  0.0121 &  0.0021 &     2.0 &     1.1 &  0.064 &  0.061 &  0.068 &  0.295\\
 0807 &    0.0 & 30.101 &  0.0630 & 182.8 & -0.0153 & -0.0385 &   -85.4 &  -183.3 & 20.963 &  5.784 & 21.898 &  6.906\\
          &        &  0.063 &  0.0170 &  43.7 &  0.0051 &  0.0096 &    14.8 &    11.8 &  0.308 &  1.900 &  0.315 &  2.765\\
          &    0.1 & 30.103 & -0.0632 & 182.2 &  0.0164 & -0.0383 &    88.7 &  -180.5 & 20.958 &  5.755 & 21.894 &  6.859\\
          &        & -0.063 &  0.0171 &  43.4 &  0.0053 &  0.0095 &    14.7 &    11.8 &  0.306 &  1.884 &  0.314 &  2.745\\
          &    0.6 & 30.059 &  0.0626 & 183.2 & -0.0921 & -0.0483 &   -81.5 &   -13.0 & 20.969 &  5.831 & 21.910 &  6.824\\
          &        &  0.063 &  0.0170 &  44.0 &  0.0248 &  0.0124 &     3.0 &     2.5 &  0.309 &  1.926 &  0.318 &  2.790\\
          &    0.8 & 30.069 & -0.0636 & 180.5 &  0.0947 & -0.0464 &    80.7 &   -10.8 & 20.951 &  5.717 & 21.890 &  6.612\\
          &        & -0.064 &  0.0172 &  43.0 &  0.0253 &  0.0118 &     3.0 &     2.4 &  0.308 &  1.881 &  0.312 &  2.697\\
 0866 &    2.7 & 14.116 &  0.4062 &  17.3 &  0.1300 &  0.0254 &   739.8 &   172.9 & 17.523 &  0.211 & 17.813 &  0.450\\
          &        &  0.406 &  0.0383 &   1.0 &  0.0284 &  0.0226 &   176.7 &    84.4 &  0.142 &  0.158 &  0.172 &  0.228\\
          &    2.7 & 14.115 & -0.4062 &  17.3 & -0.1298 &  0.0259 &  -740.2 &   175.9 & 17.523 &  0.211 & 17.814 &  0.450\\
          &        & -0.406 &  0.0383 &   1.0 &  0.0283 &  0.0227 &   177.0 &    84.1 &  0.142 &  0.158 &  0.172 &  0.228\\
          &    0.1 & 14.119 &  0.4183 &  17.1 & -1.1858 & -0.1739 &   -84.0 &    16.2 & 17.478 &  0.162 & 17.801 &  0.429\\
          &        &  0.418 &  0.0405 &   1.0 &  0.1158 &  0.0220 &     3.4 &     1.5 &  0.147 &  0.157 &  0.175 &  0.229\\
          &    0.0 & 14.121 & -0.4186 &  17.1 &  1.1861 & -0.1788 &    83.2 &    15.9 & 17.477 &  0.161 & 17.799 &  0.427\\
          &        & -0.419 &  0.0406 &   1.0 &  0.1159 &  0.0223 &     3.4 &     1.5 &  0.147 &  0.157 &  0.175 &  0.229\\
 0874 &   19.7 & 45.665 &  0.1852 &  25.5 & -0.1028 &  0.0212 &  -635.1 &   159.2 & 15.908 &  0.023 & 17.350 &  1.387\\
          &        &  0.185 &  0.0007 &   0.1 &  0.0052 &  0.0040 &    29.9 &    26.9 &  0.005 &  0.004 &  0.059 &  0.247\\
          &    3.1 & 45.668 & -0.1855 &  25.4 &  0.1074 &  0.0092 &   629.0 &    82.1 & 15.906 &  0.021 & 17.377 &  1.456\\
          &        & -0.186 &  0.0007 &   0.1 &  0.0049 &  0.0039 &    27.9 &    23.9 &  0.005 &  0.004 &  0.059 &  0.252\\
          &   36.4 & 45.662 &  0.1850 &  25.5 & -0.2097 &  0.0225 &  -321.0 &    62.7 & 15.910 &  0.024 & 17.382 &  1.494\\
          &        &  0.185 &  0.0007 &   0.1 &  0.0049 &  0.0038 &     7.5 &     5.7 &  0.005 &  0.004 &  0.058 &  0.253\\
          &    0.0 & 45.669 & -0.1857 &  25.4 &  0.2132 & -0.0002 &   319.3 &    28.0 & 15.905 &  0.020 & 17.374 &  1.434\\
          &        & -0.186 &  0.0007 &   0.1 &  0.0051 &  0.0038 &     7.5 &     5.7 &  0.005 &  0.004 &  0.059 &  0.252\\
 0944 &    0.6 & 12.751 &  0.2742 &   9.9 &  0.0801 & -0.1700 &   397.2 &  -815.4 & 15.668 & -0.085 & 15.648 &  0.099\\
          &        &  0.274 &  0.0077 &   0.2 &  0.0110 &  0.0113 &    75.2 &    21.8 &  0.037 &  0.031 &  0.063 &  0.065\\
          &    0.8 & 12.750 & -0.2741 &   9.9 & -0.1071 & -0.1605 &  -505.4 &  -727.6 & 15.668 & -0.084 & 15.638 &  0.089\\
          &        & -0.274 &  0.0077 &   0.2 &  0.0103 &  0.0129 &    72.7 &    26.1 &  0.036 &  0.031 &  0.064 &  0.065\\
          &    1.7 & 12.749 &  0.2741 &   9.9 & -0.7638 & -0.1578 &  -220.9 &   -16.7 & 15.669 & -0.084 & 15.605 &  0.056\\
          &        &  0.274 &  0.0077 &   0.2 &  0.0245 &  0.0122 &     2.7 &     4.6 &  0.036 &  0.031 &  0.065 &  0.064\\
          &    0.0 & 12.752 & -0.2744 &   9.9 &  0.7075 & -0.2602 &   217.6 &   -51.5 & 15.667 & -0.085 & 15.700 &  0.151\\
          &        & -0.274 &  0.0077 &   0.2 &  0.0231 &  0.0105 &     2.1 &     4.7 &  0.037 &  0.031 &  0.061 &  0.066\\
 0979 &    0.0 & 13.737 &  0.1064 &   8.9 & -0.0076 & -0.0330 & -1291.2 & -5550.2 & 17.460 &  0.108 & 18.925 & -0.162\\
          &        &  0.106 &  0.0044 &   0.3 &  0.0046 &  0.0166 &   941.7 &  2774.6 &  0.049 &  0.050 &  0.088 &  0.076\\
          &    0.0 & 13.737 & -0.1064 &   8.9 & -0.0114 & -0.0326 & -1849.3 & -5277.8 & 17.460 &  0.108 & 18.925 & -0.162\\
          &        & -0.106 &  0.0044 &   0.3 &  0.0070 &  0.0160 &   988.5 &  2699.6 &  0.049 &  0.050 &  0.088 &  0.076\\
          &    0.0 & 13.737 &  0.1064 &   8.9 & -0.2711 &  0.0365 &  -700.9 &   122.9 & 17.461 &  0.109 & 18.905 & -0.173\\
          &        &  0.106 &  0.0044 &   0.3 &  0.0116 &  0.0170 &    10.8 &    45.3 &  0.049 &  0.050 &  0.088 &  0.076\\
          &    0.2 & 13.738 & -0.1065 &   8.9 &  0.2440 & -0.1105 &   660.7 &  -270.1 & 17.460 &  0.107 & 18.948 & -0.150\\
          &        & -0.106 &  0.0044 &   0.3 &  0.0113 &  0.0162 &    27.2 &    37.6 &  0.049 &  0.050 &  0.087 &  0.077\\
 1021 &    0.6 & 23.214 &  0.0604 &  13.2 &  0.0729 &  0.0359 &  1451.7 &   744.3 & 18.072 & -0.028 & 17.998 &  2.126\\
          &        &  0.060 &  0.0027 &   0.4 &  0.0051 &  0.0031 &    86.9 &    66.1 &  0.047 &  0.042 &  0.056 &  0.172\\
          &    1.3 & 23.214 & -0.0604 &  13.2 & -0.0680 &  0.0458 & -1331.6 &   926.6 & 18.072 & -0.028 & 17.992 &  2.111\\
          &        & -0.060 &  0.0027 &   0.4 &  0.0049 &  0.0035 &    81.5 &    72.2 &  0.047 &  0.042 &  0.057 &  0.172\\
          &    1.8 & 23.214 &  0.0604 &  13.2 & -0.1940 &  0.0513 &  -635.1 &   196.8 & 18.073 & -0.028 & 17.987 &  2.098\\
          &        &  0.060 &  0.0027 &   0.4 &  0.0088 &  0.0035 &    16.9 &     9.0 &  0.047 &  0.042 &  0.057 &  0.172\\
          &    0.0 & 23.214 & -0.0604 &  13.2 &  0.1973 &  0.0243 &   656.1 &   110.4 & 18.072 & -0.028 & 18.004 &  2.141\\
          &        & -0.060 &  0.0027 &   0.4 &  0.0089 &  0.0029 &    16.9 &     8.9 &  0.047 &  0.042 &  0.056 &  0.173\\
 1147 &    0.5 & 37.488 &  0.7191 &   7.7 & -0.1197 & -0.0554 & -1554.7 &  -689.8 & 15.487 &  0.121 & 15.834 &  0.756\\
          &        &  0.719 &  0.0854 &   0.6 &  0.0254 &  0.0112 &   225.8 &   213.7 &  0.227 &  0.234 &  0.240 &  0.415\\
          &    0.6 & 37.489 & -0.7192 &   7.7 &  0.1202 & -0.0576 &  1523.3 &  -703.0 & 15.487 &  0.120 & 15.833 &  0.751\\
          &        & -0.719 &  0.0854 &   0.6 &  0.0256 &  0.0115 &   220.8 &   213.8 &  0.227 &  0.234 &  0.240 &  0.414\\
          &    0.0 & 37.484 &  0.7192 &   7.7 & -1.1089 & -0.0929 &  -204.9 &    11.3 & 15.487 &  0.120 & 15.833 &  0.767\\
          &        &  0.719 &  0.0854 &   0.6 &  0.1326 &  0.0146 &     9.5 &     2.2 &  0.227 &  0.234 &  0.240 &  0.416\\
          &    0.8 & 37.491 & -0.7206 &   7.7 &  1.0986 & -0.1023 &   201.6 &     9.2 & 15.483 &  0.116 & 15.829 &  0.726\\
          &        & -0.721 &  0.0858 &   0.6 &  0.1314 &  0.0150 &     9.5 &     2.2 &  0.228 &  0.234 &  0.241 &  0.412\\
\enddata
\tablecomments{Fit parameters for the ensemble of 20 out 
the 21 events discussed in the text. 
For the analysis of OGLE-2014-BLG-1049 we refer to the text 
and Figure~\ref{fig:1049}.
For each event we report the 4 solutions in the order $-+, --, ++, +-$. 
The lightcurves and the the ellipses for each solution
for the heliocentric velocity and the parallax are show in Figure~\ref{fig:lc}.}
\tablenotetext{a}{$f$ indicates the ratio of blend to source flux.}
\tablenotetext{b}{Instrumental magnitude.}
\end{deluxetable}


\vfil\eject

\begin{subfigures}

\begin{figure}
\label{fig:lc}
\epsscale{.90}
\plotone{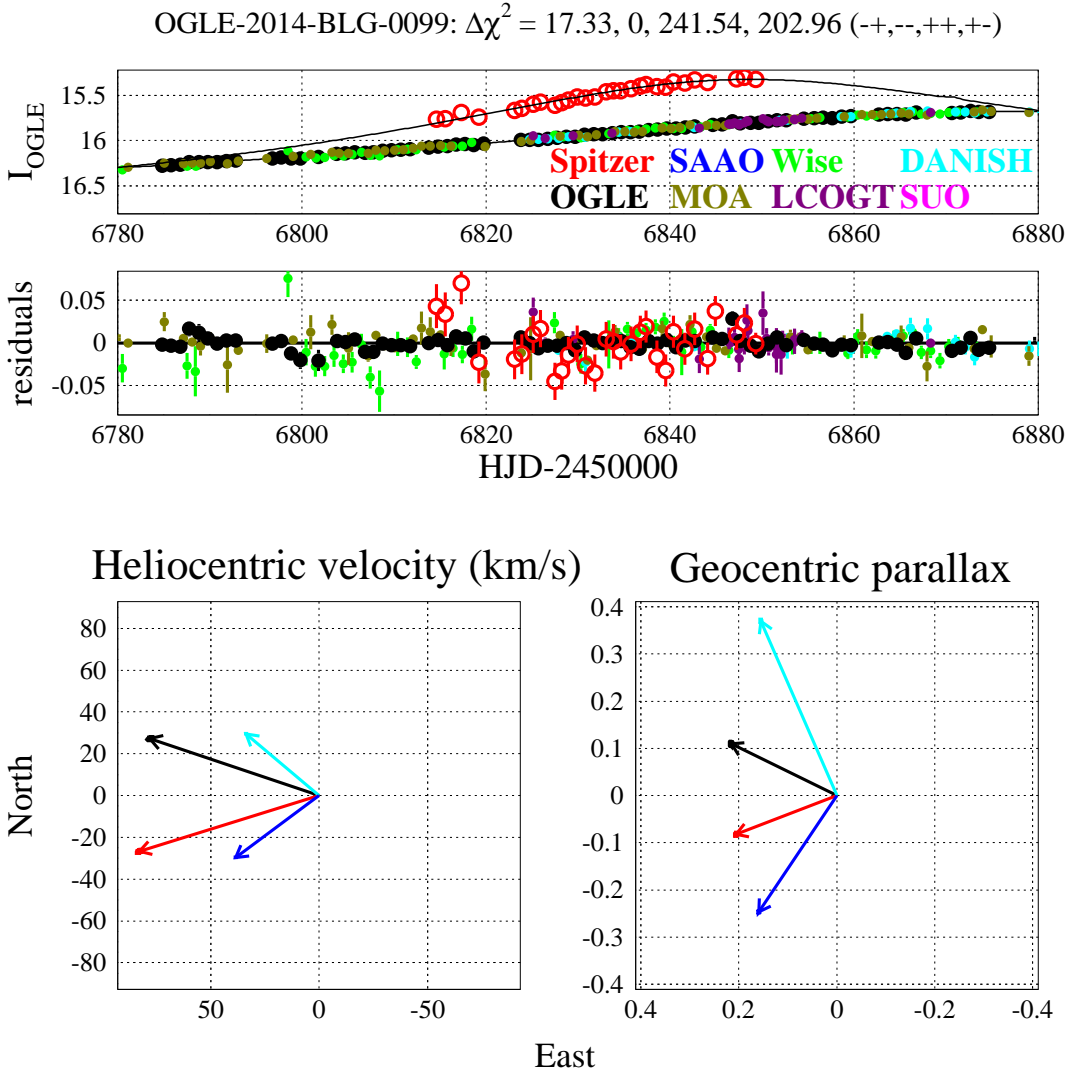}
\caption{OGLE-2014-BLG-0099. 
Top: the lightcurve data together with the 
Spitzer and the ground-based best fit models.
Second panel from top: residual light curve.
In both panels the Spitzer and the ground-based data are
shown as empty and filled circles, respectively.
For purposes of display, all the data set
are binned with 1 point per epoch.
The color codes are indicated in the top panel:
red, black, blue, olive green, green and purple
for Spitzer, OGLE, SAAO (PLANET),
MOA, Wise, LCOGT (RoboNet, the details of the different
telescopes of the network used is given in Table~\ref{tab:evt_param})
and for the MiNDSTEp collaboration the Danish (cyan)
and the Salerno University Telescope (magenta).
In the two bottom panels
we show the projected heliocentric velocity $\tilde\bv_\hel$ (left)
and the geocentric parallax vectors $\bpi_{\e,\geo}$ and
ellipse errors (which can however be too small to be seen),
in the North-East equatorial frame, as given in Table~\ref{tab:evt}.
The values of the $\Delta\chi^2$, 
as reported in the title, are color-coded
as black, red, cyan and blue, from the best solution to the worst.
}
\end{figure}
\clearpage

\begin{figure}
\plotone{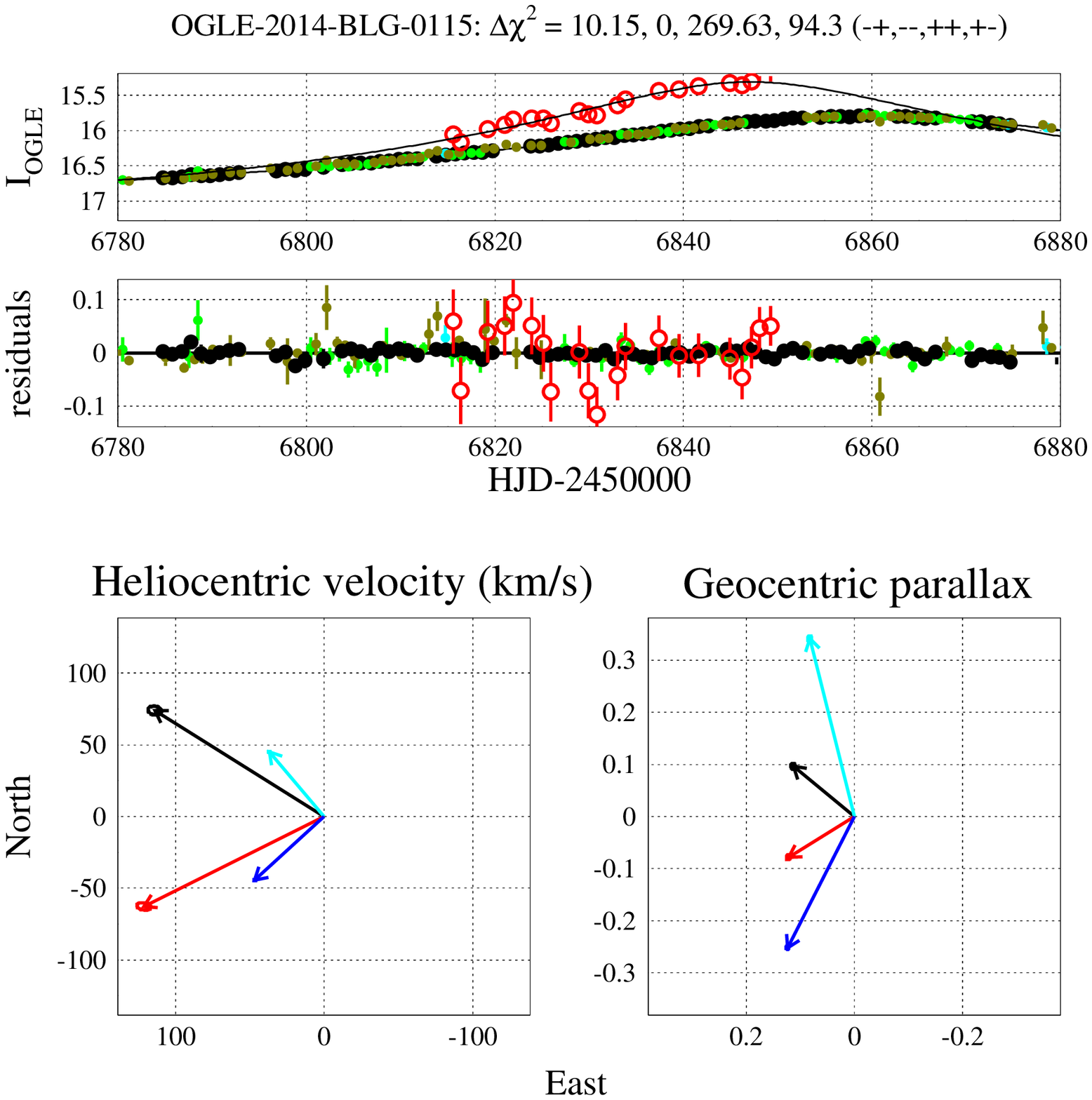}
\caption{OGLE-2014-BLG-0115. Panels and symbols as in Figure~\ref{fig:lc}a.}
\end{figure}
\clearpage

\begin{figure}
\plotone{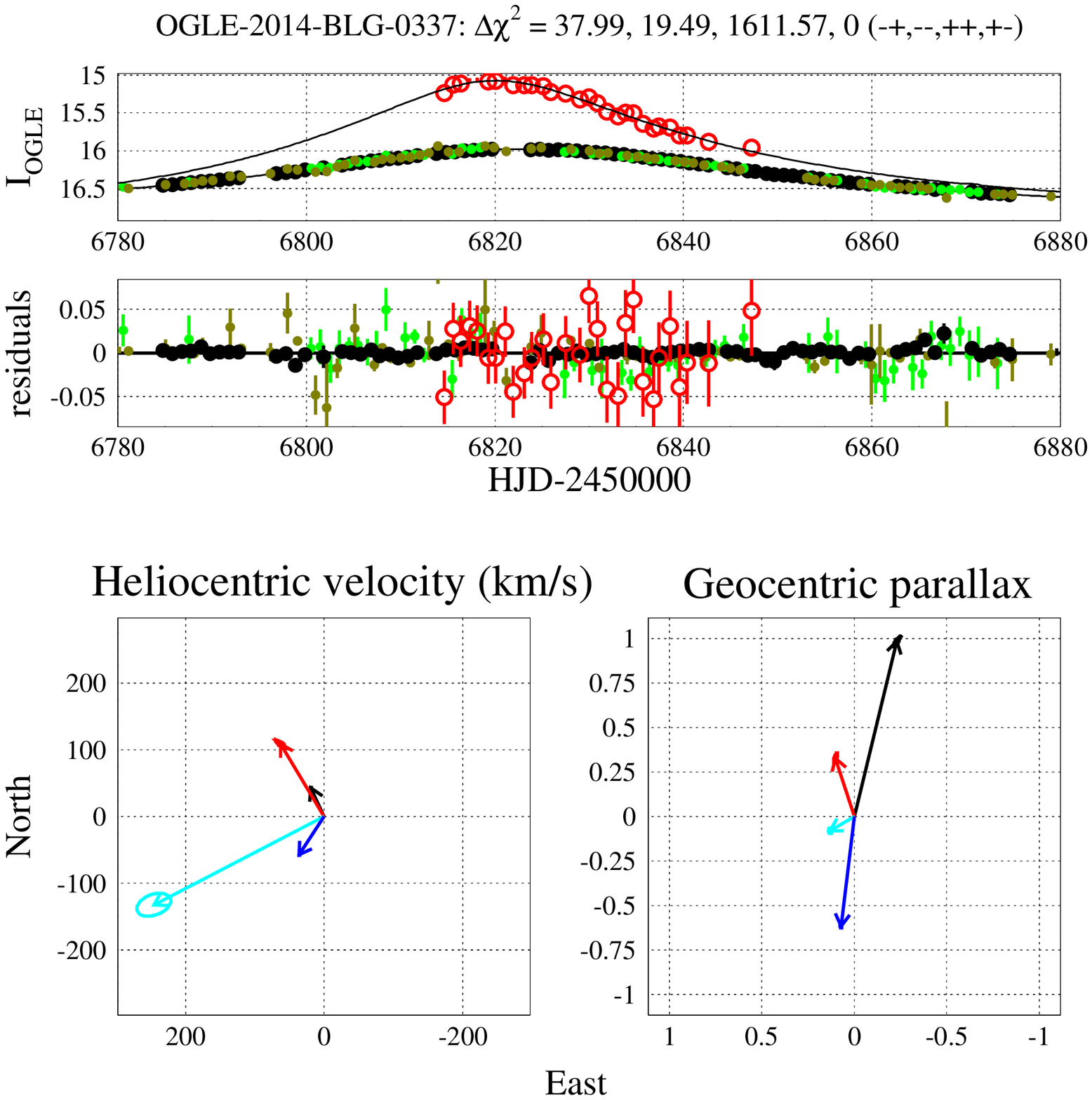}
\caption{OGLE-2014-BLG-0337. Panels and symbols as in Figure~\ref{fig:lc}a.}
\end{figure}
\clearpage

\begin{figure}
\plotone{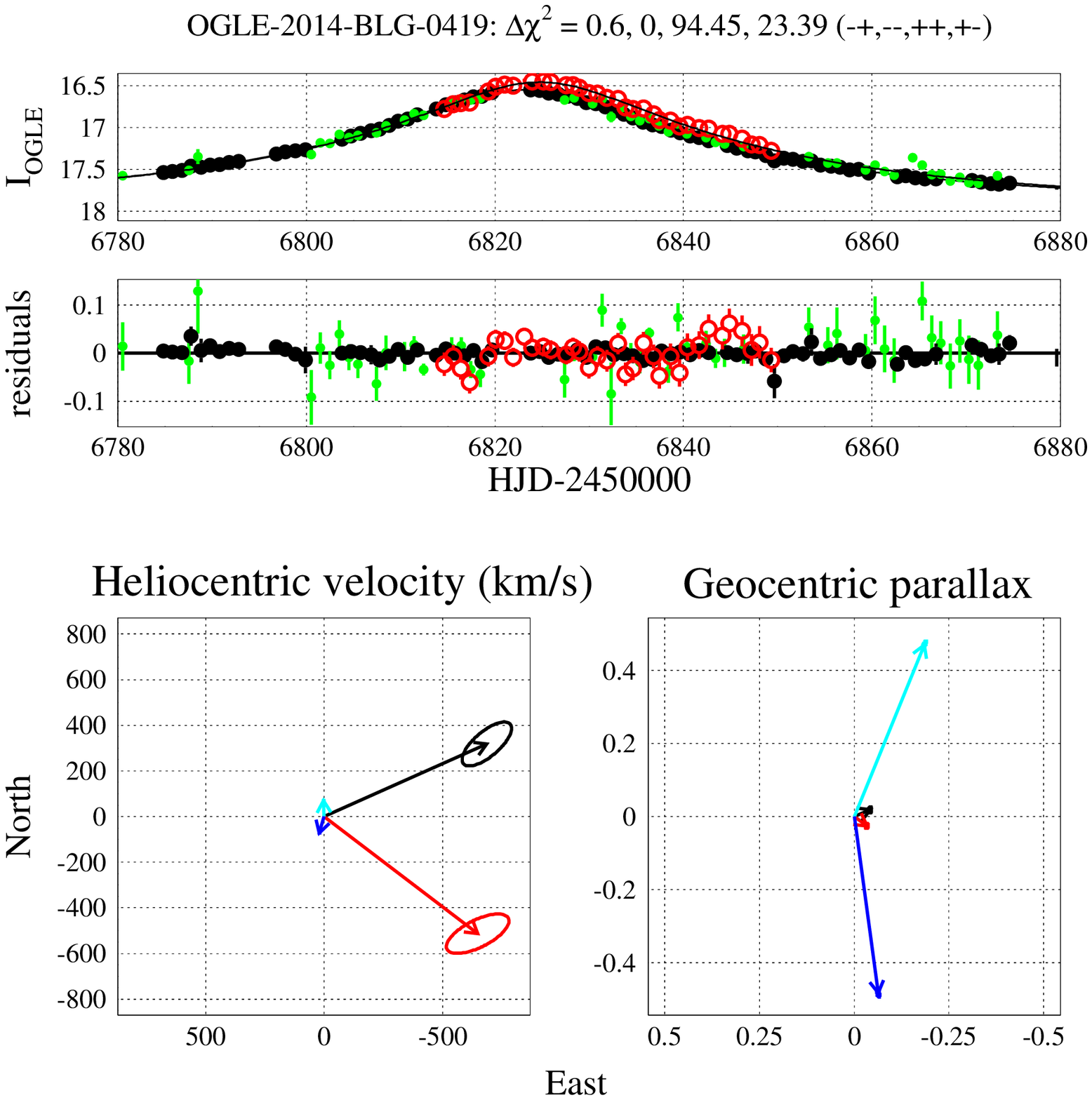}
\caption{OGLE-2014-BLG-0419. Panels and symbols as in Figure~\ref{fig:lc}a.}
\end{figure}
\clearpage

\begin{figure}
\plotone{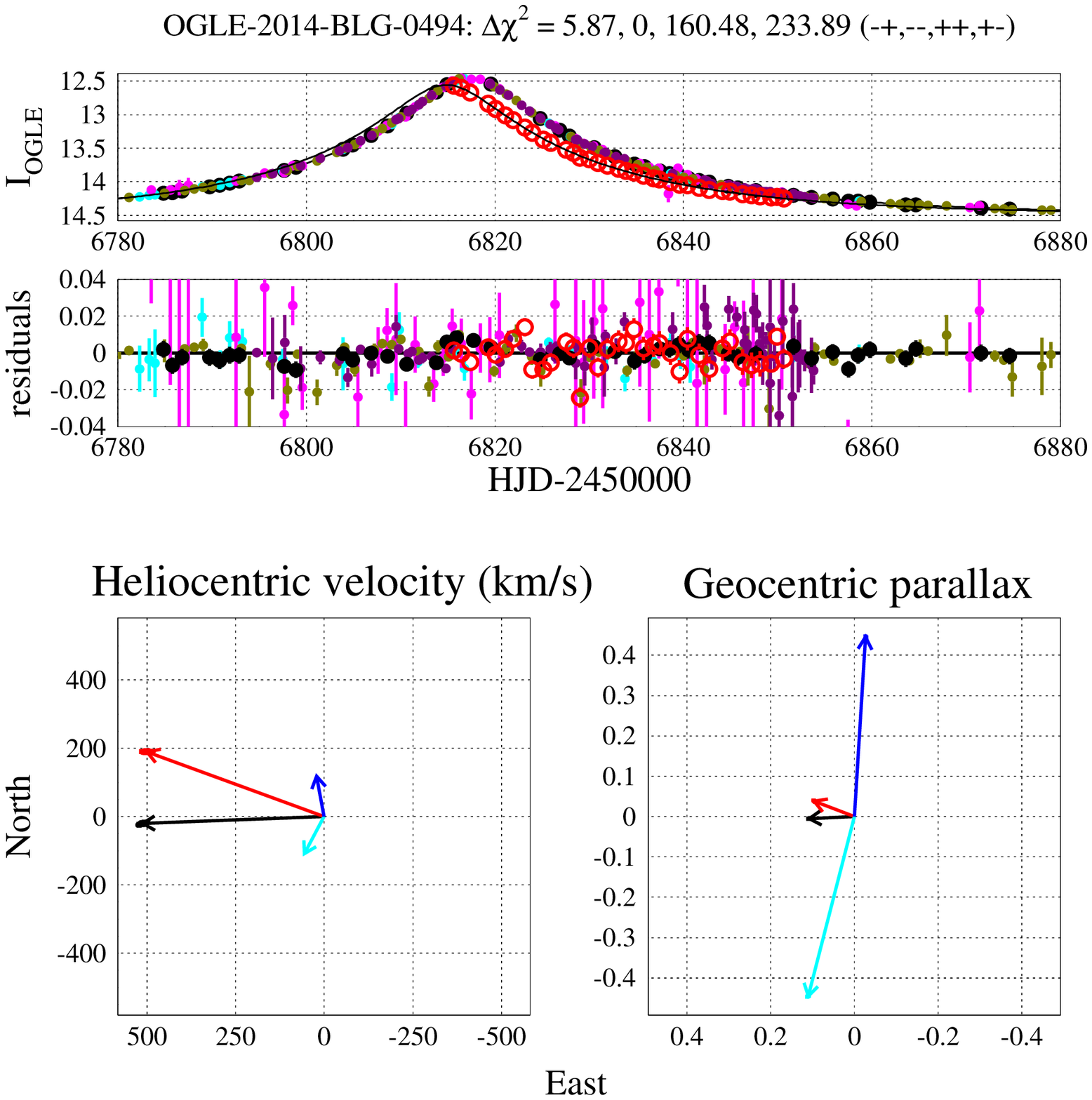}
\caption{OGLE-2014-BLG-0494. Panels and symbols as in Figure~\ref{fig:lc}a.}
\end{figure}
\clearpage

\begin{figure}
\plotone{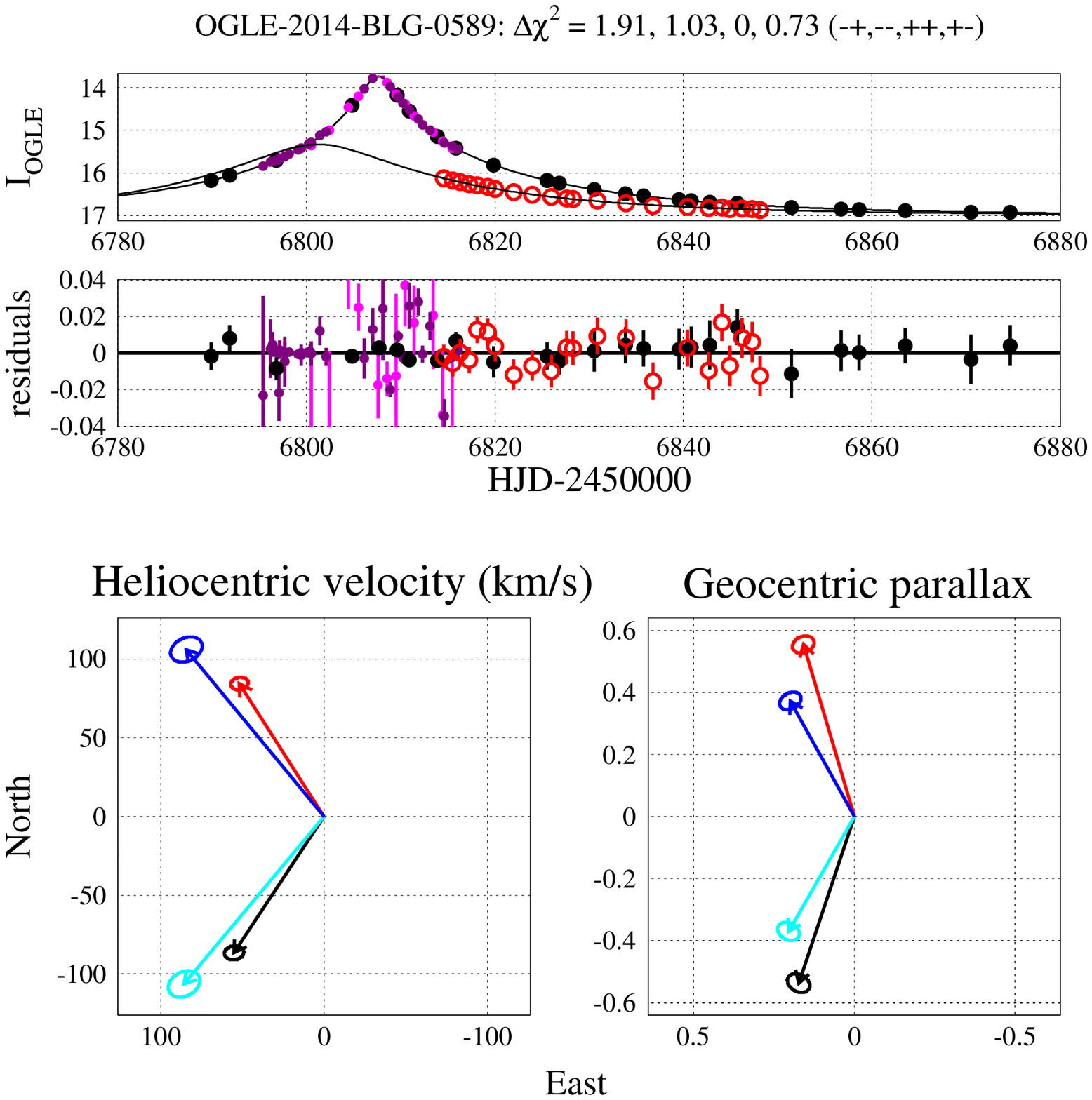}
\caption{OGLE-2014-BLG-0589. Panels and symbols as in Figure~\ref{fig:lc}a.}
\end{figure}
\clearpage

\begin{figure}
\plotone{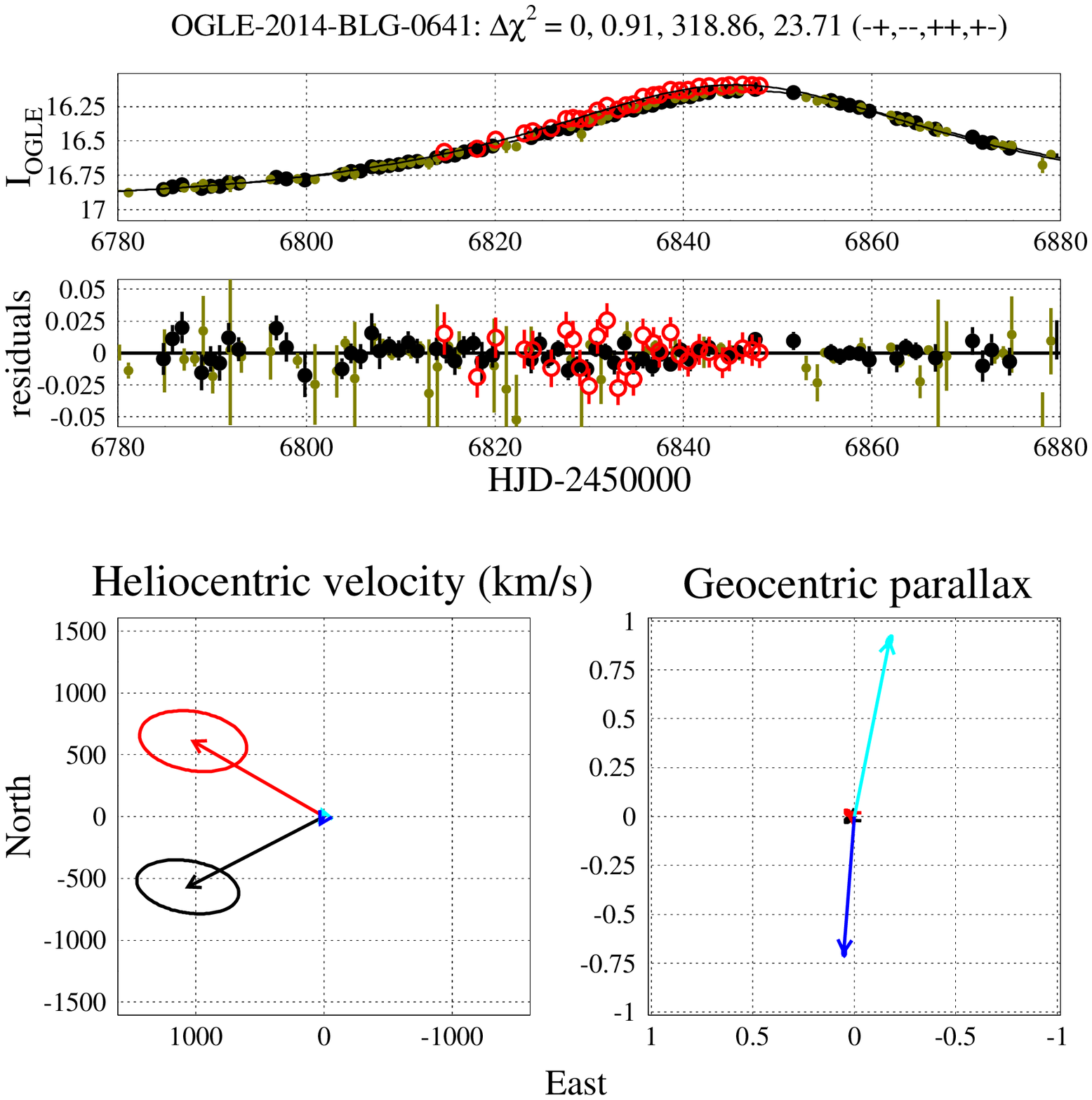}
\caption{OGLE-2014-BLG-0641. Panels and symbols as in Figure~\ref{fig:lc}a.}
\end{figure}
\clearpage

\begin{figure}
\plotone{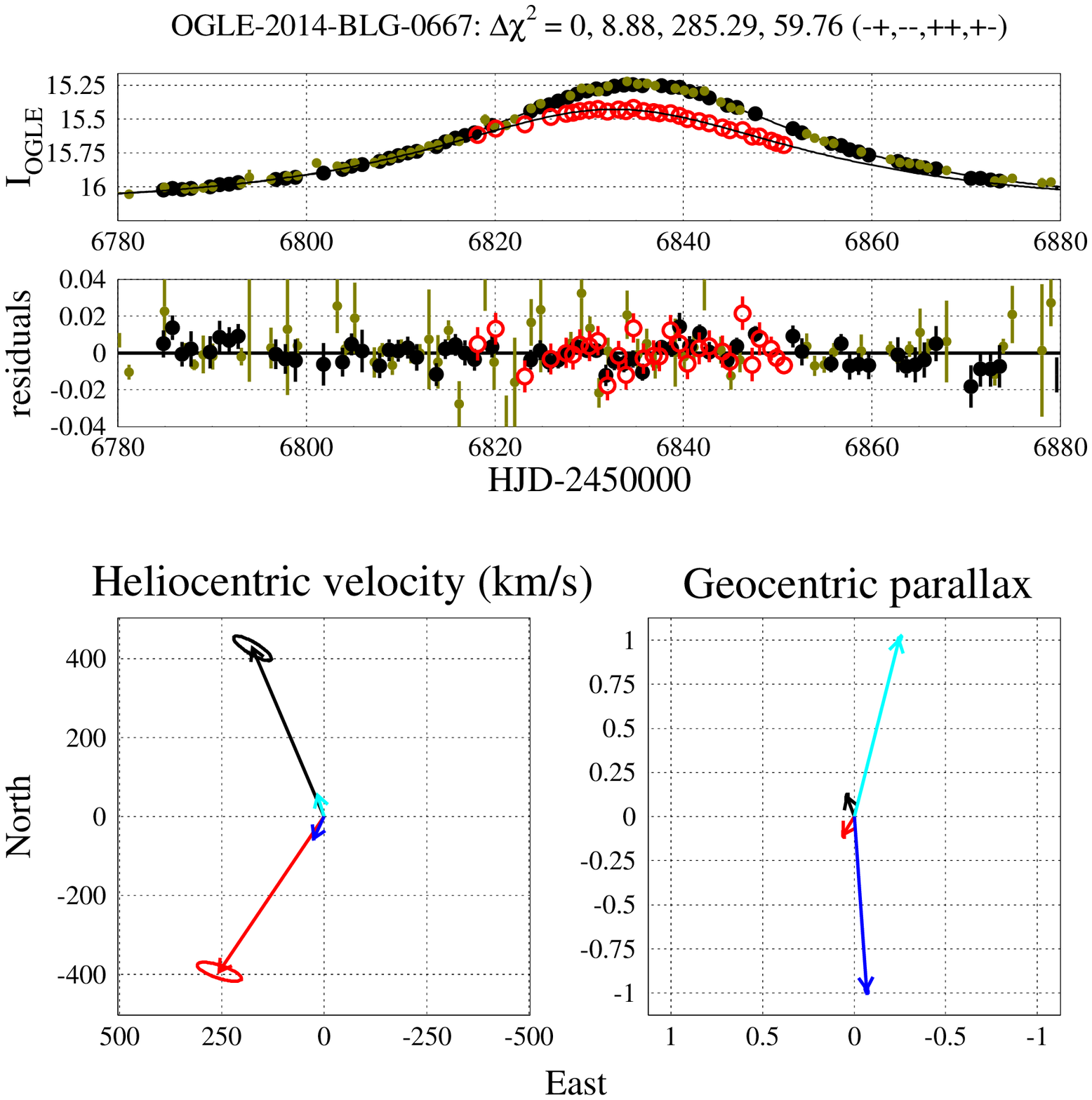}
\caption{OGLE-2014-BLG-0667. Panels and symbols as in Figure~\ref{fig:lc}a.}
\end{figure}
\clearpage

\begin{figure}
\plotone{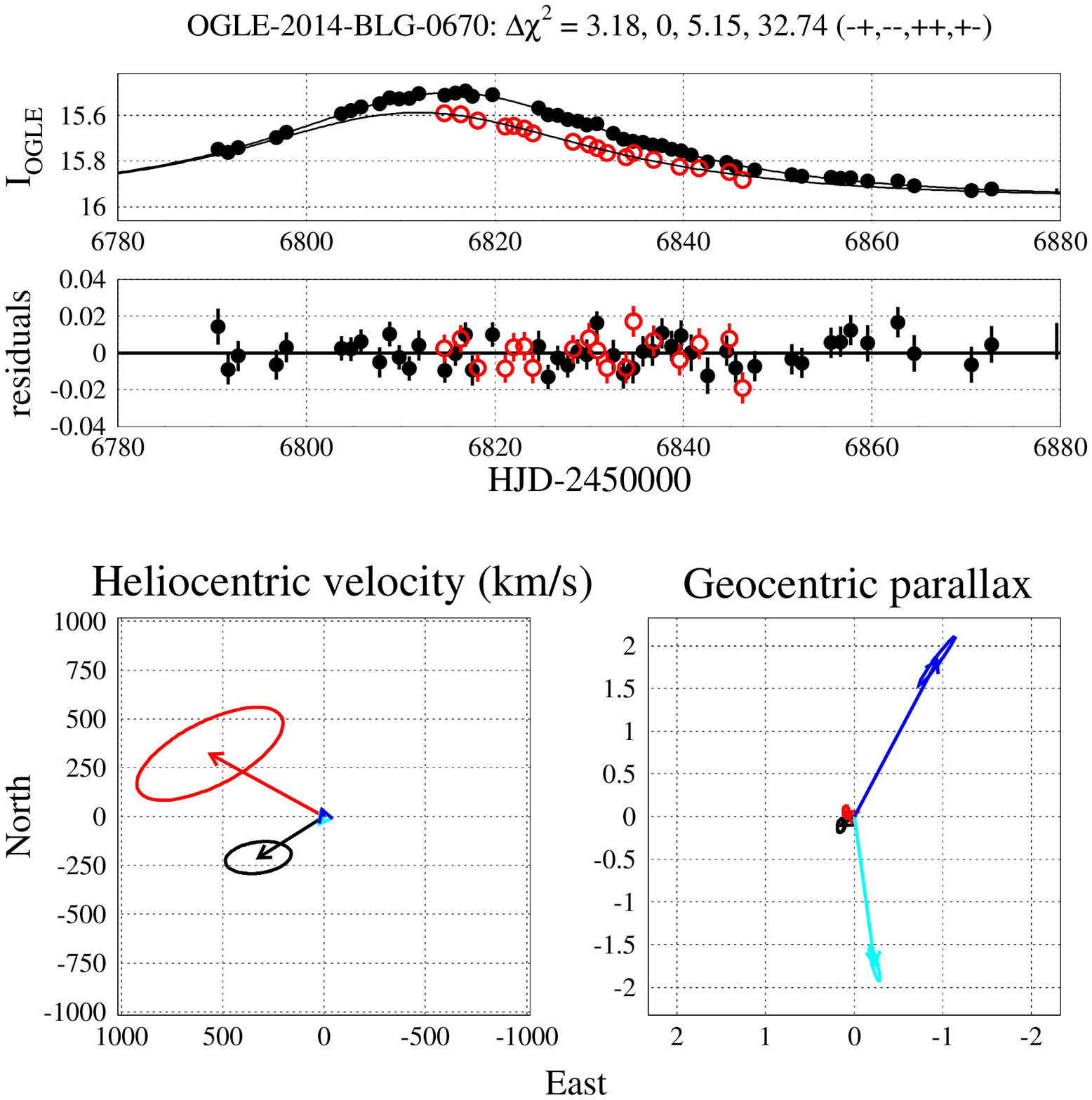}
\caption{OGLE-2014-BLG-0670. Panels and symbols as in Figure~\ref{fig:lc}a.}
\end{figure}
\clearpage

\begin{figure}
\plotone{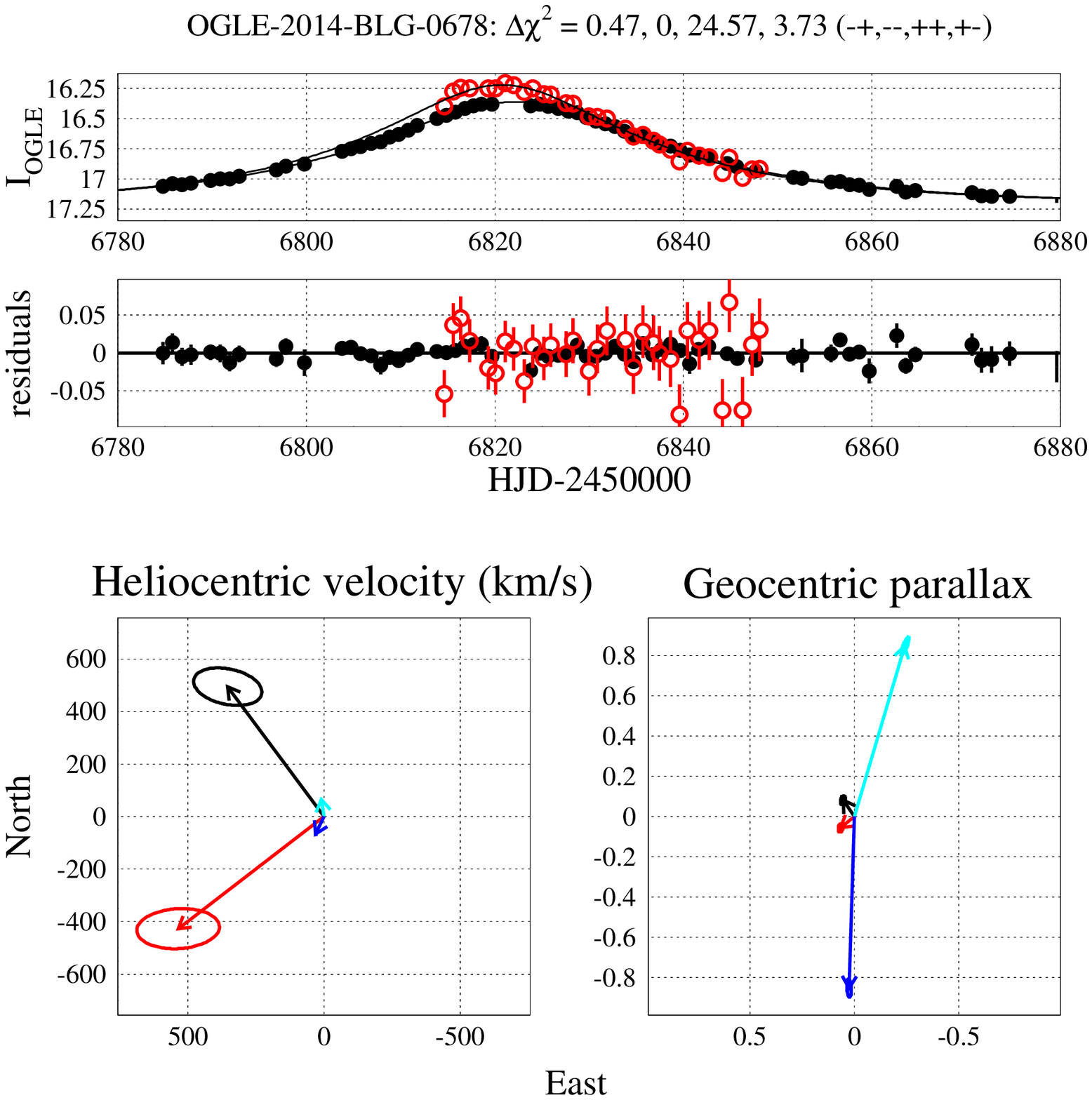}
\caption{OGLE-2014-BLG-0678. Panels and symbols as in Figure~\ref{fig:lc}a.}
\end{figure}
\clearpage

\begin{figure}
\plotone{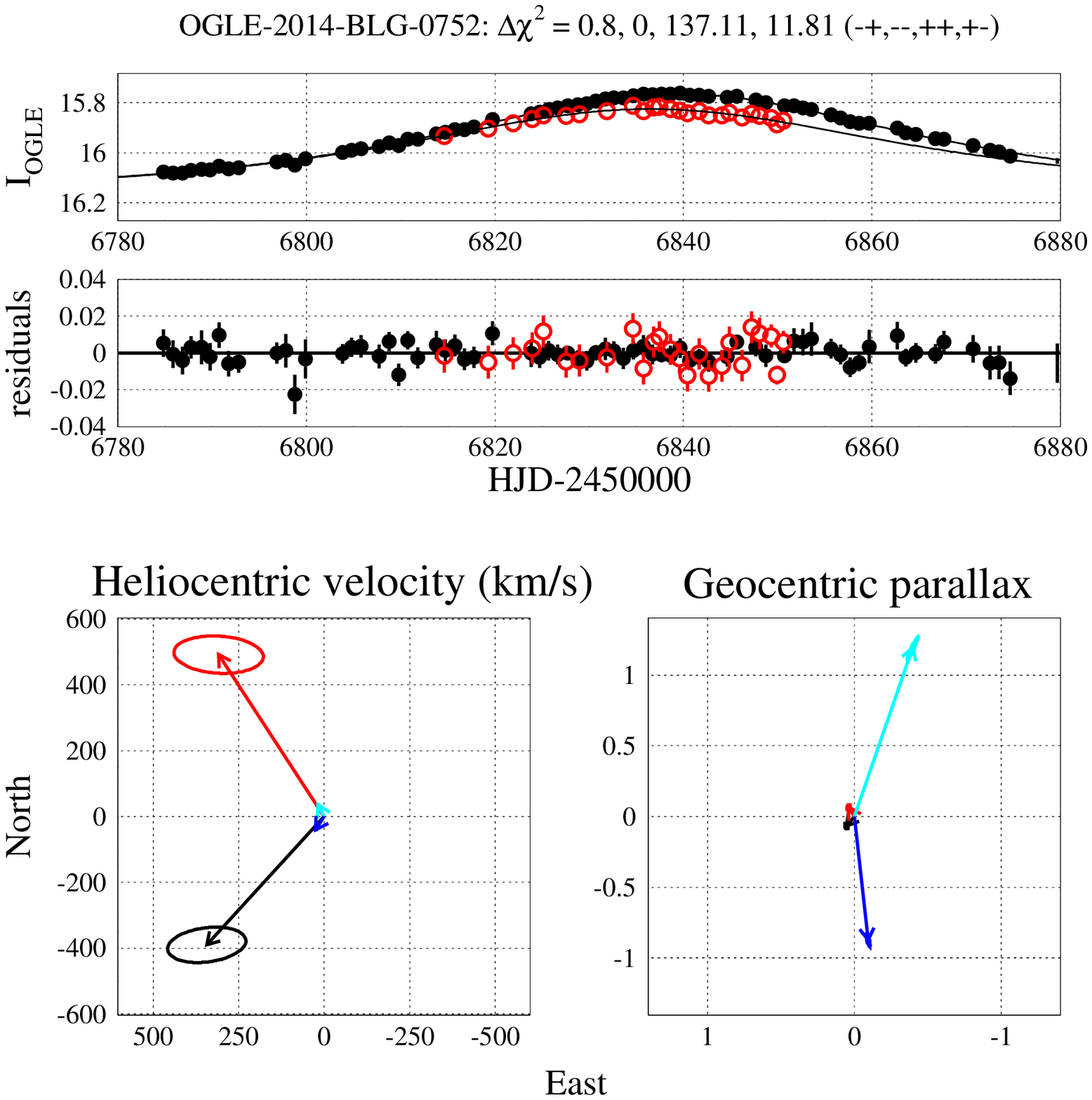}
\caption{OGLE-2014-BLG-0752. Panels and symbols as in Figure~\ref{fig:lc}a.}
\end{figure}
\clearpage

\begin{figure}
\plotone{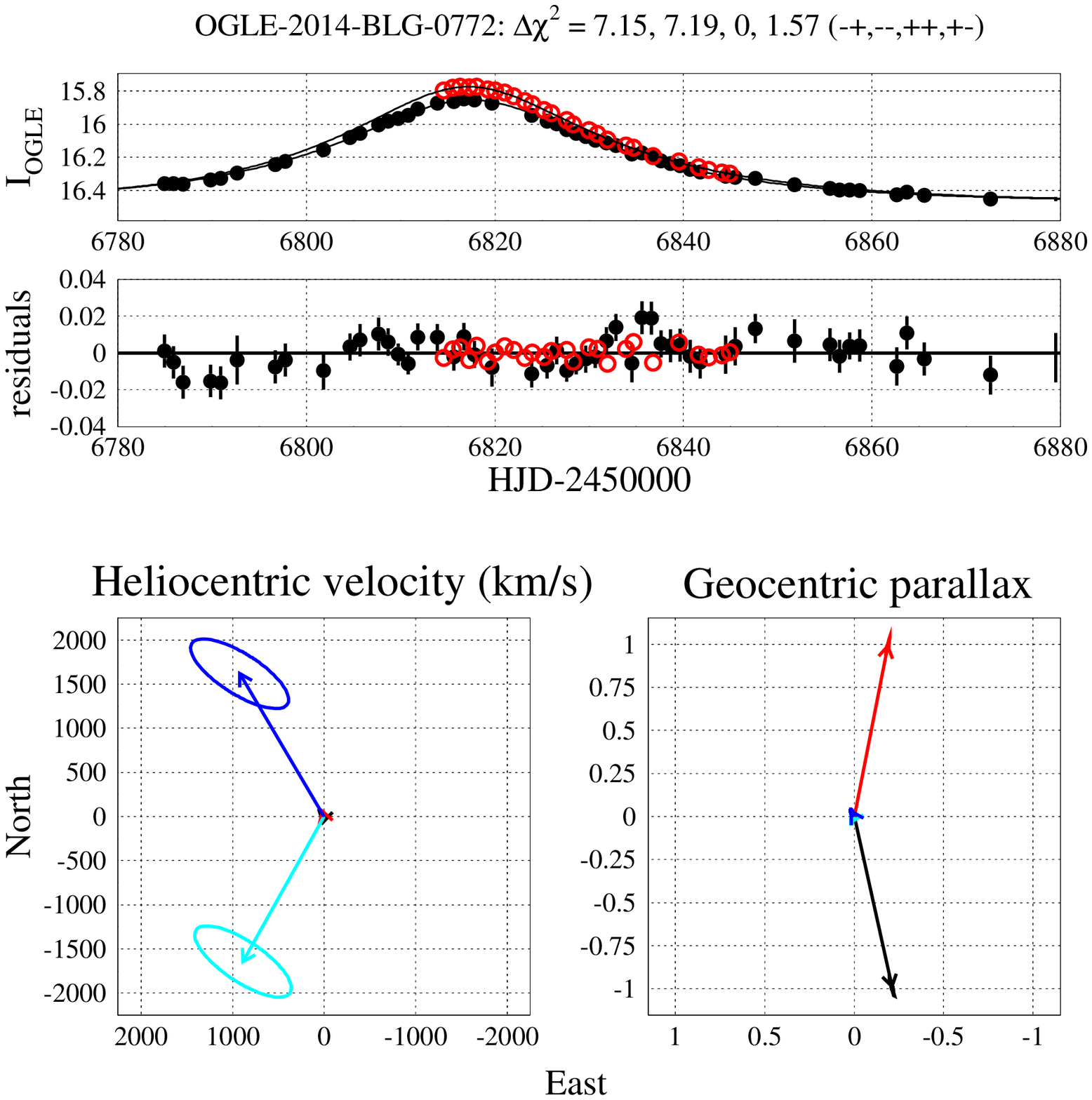}
\caption{OGLE-2014-BLG-0772. Panels and symbols as in Figure~\ref{fig:lc}a.}
\end{figure}
\clearpage

\begin{figure}
\plotone{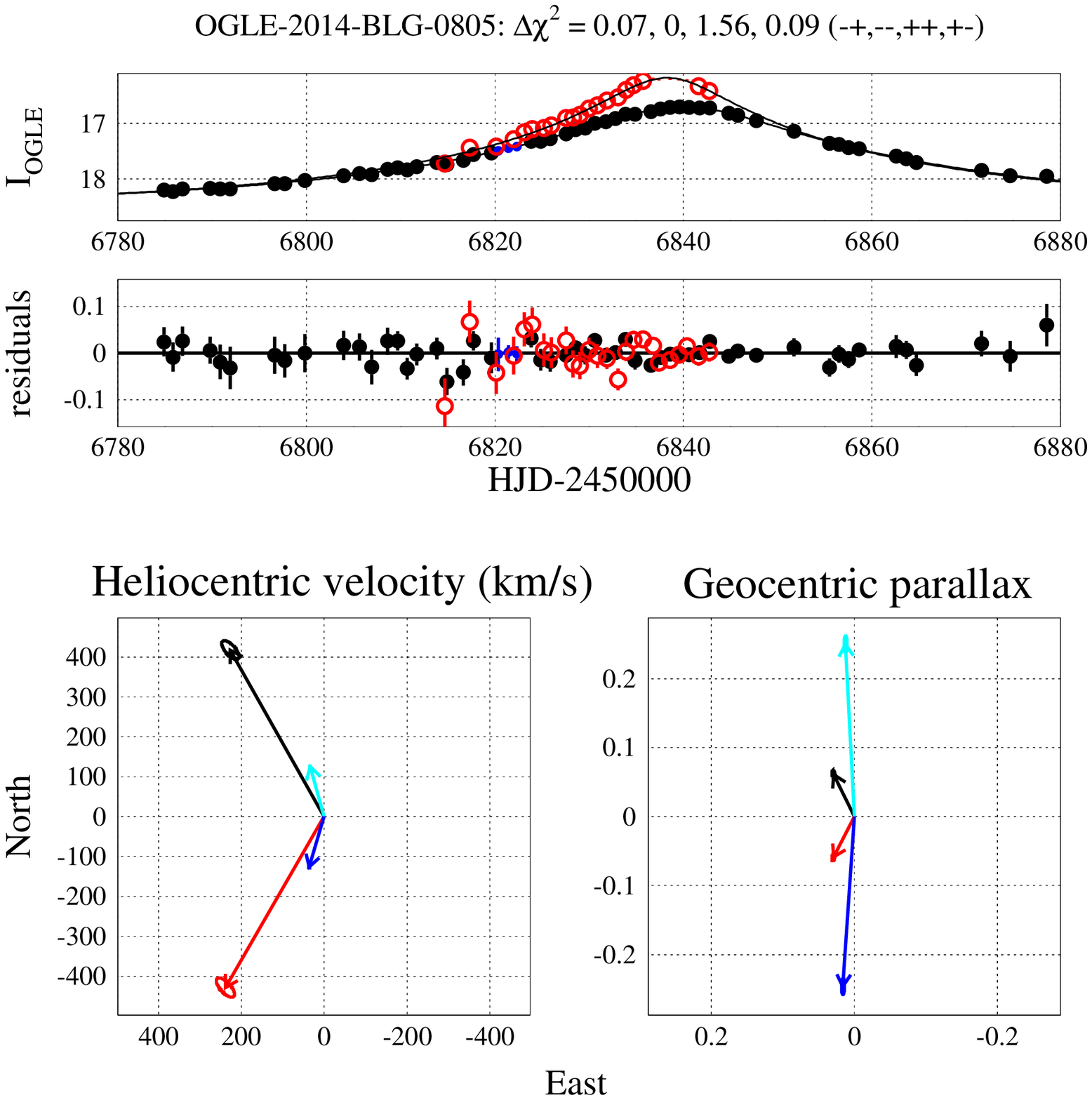}
\caption{OGLE-2014-BLG-0805. Panels and symbols as in Figure~\ref{fig:lc}a.}
\end{figure}
\clearpage

\begin{figure}
\plotone{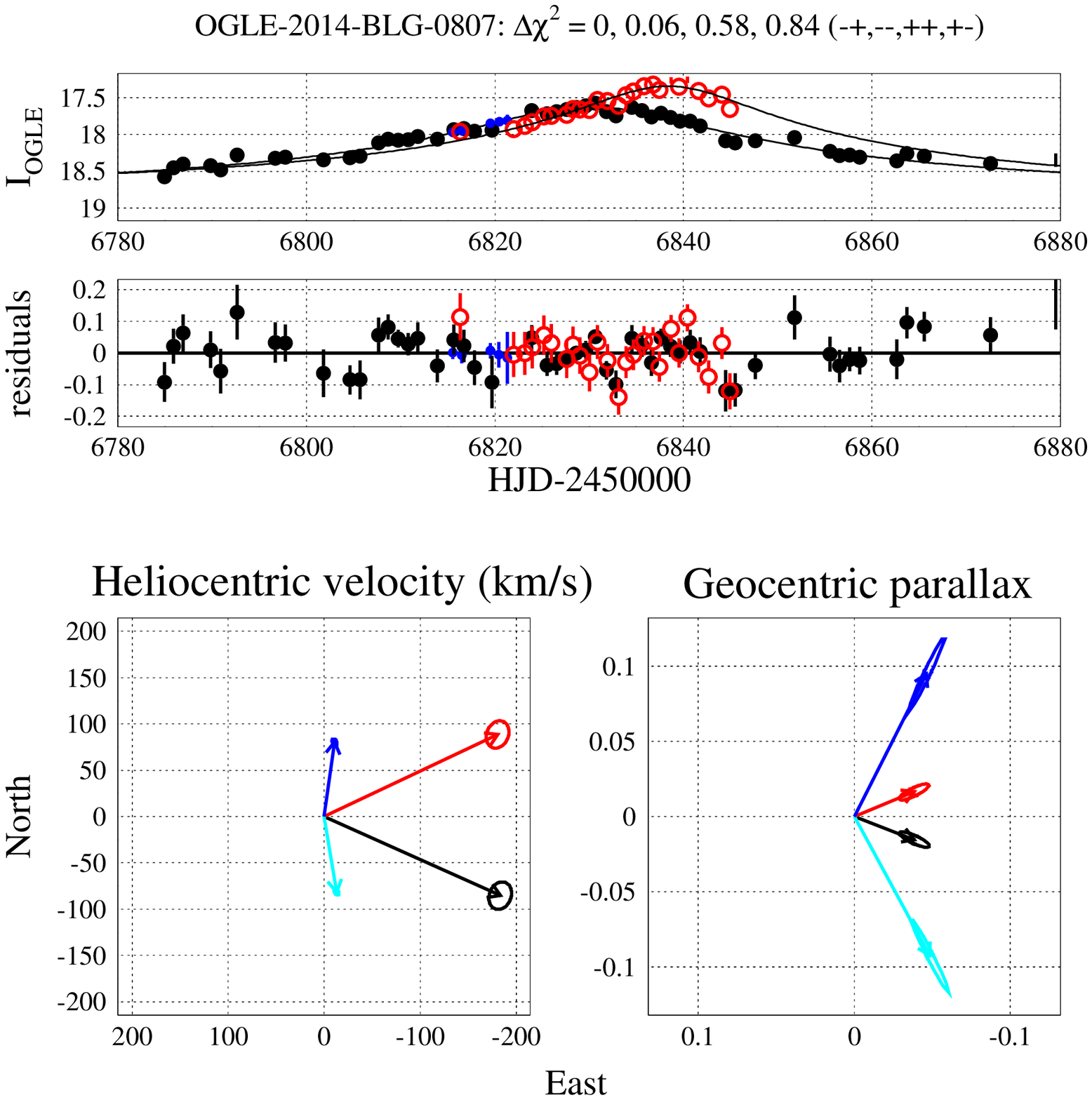}
\caption{OGLE-2014-BLG-0807. Panels and symbols as in Figure~\ref{fig:lc}a.}
\end{figure}
\clearpage

\begin{figure}
\plotone{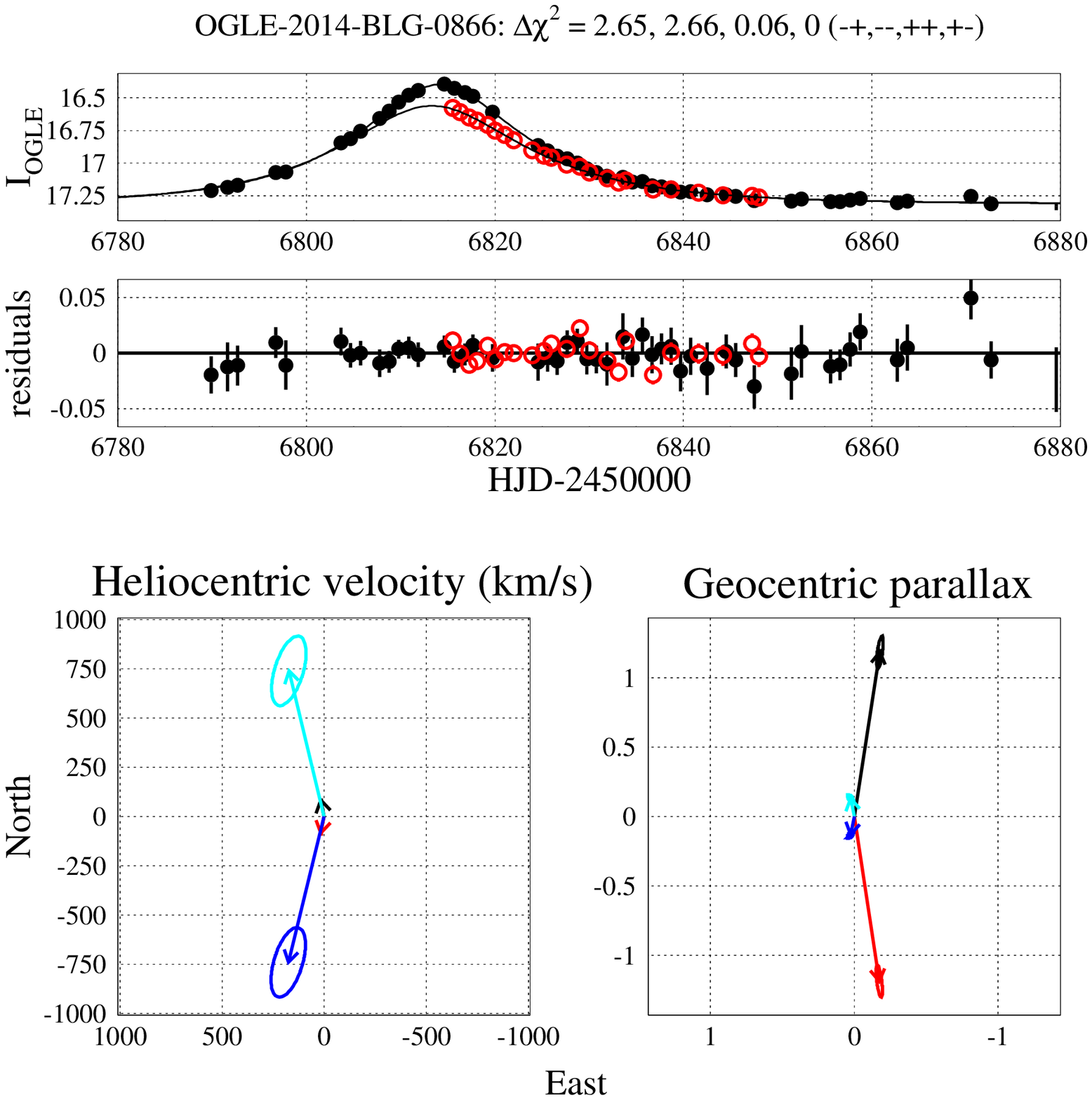}
\caption{OGLE-2014-BLG-0866. Panels and symbols as in Figure~\ref{fig:lc}a.}
\end{figure}
\clearpage

\begin{figure}
\plotone{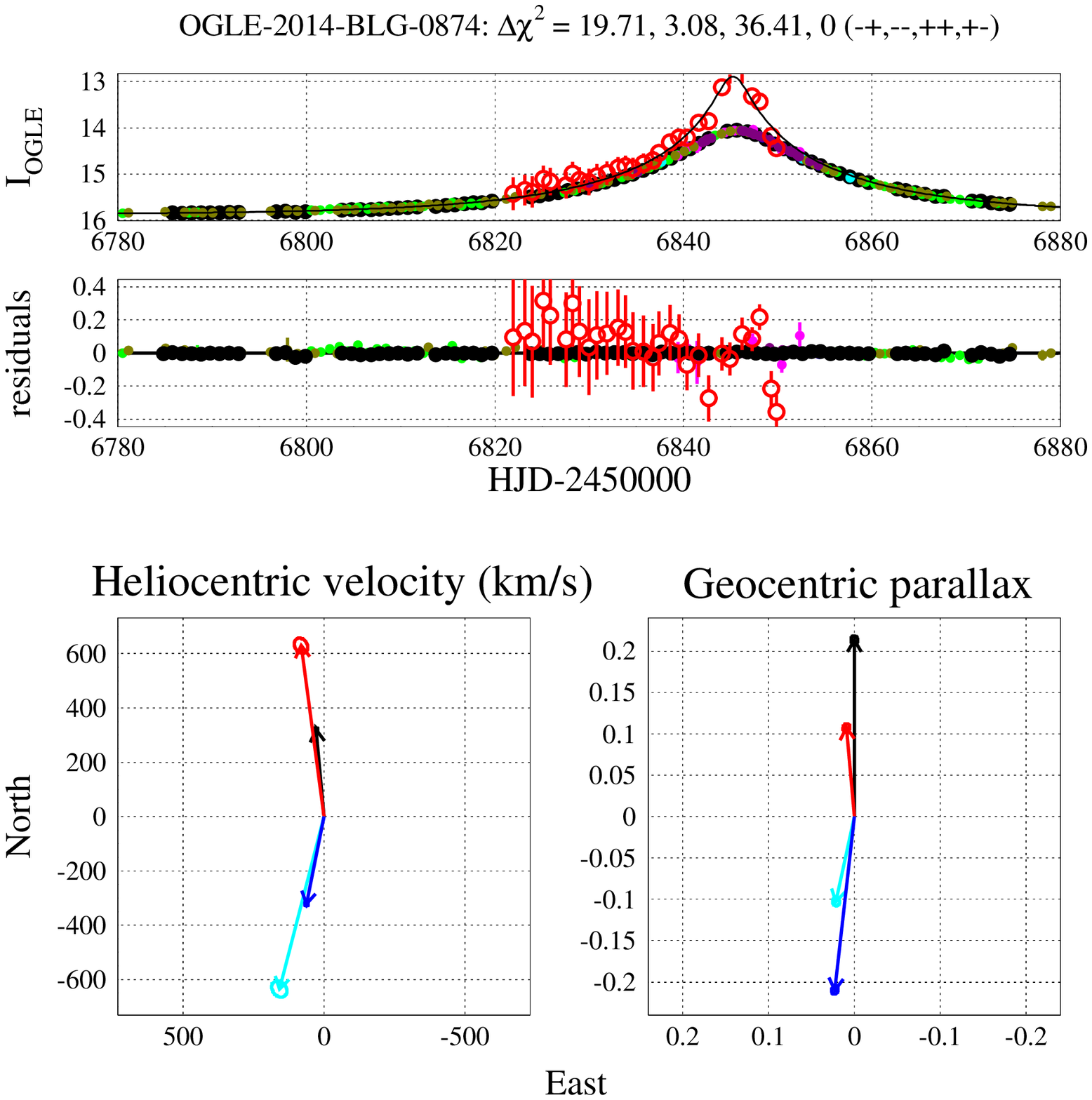}
\caption{OGLE-2014-BLG-0874. Panels and symbols as in Figure~\ref{fig:lc}a.}
\end{figure}
\clearpage

\begin{figure}
\plotone{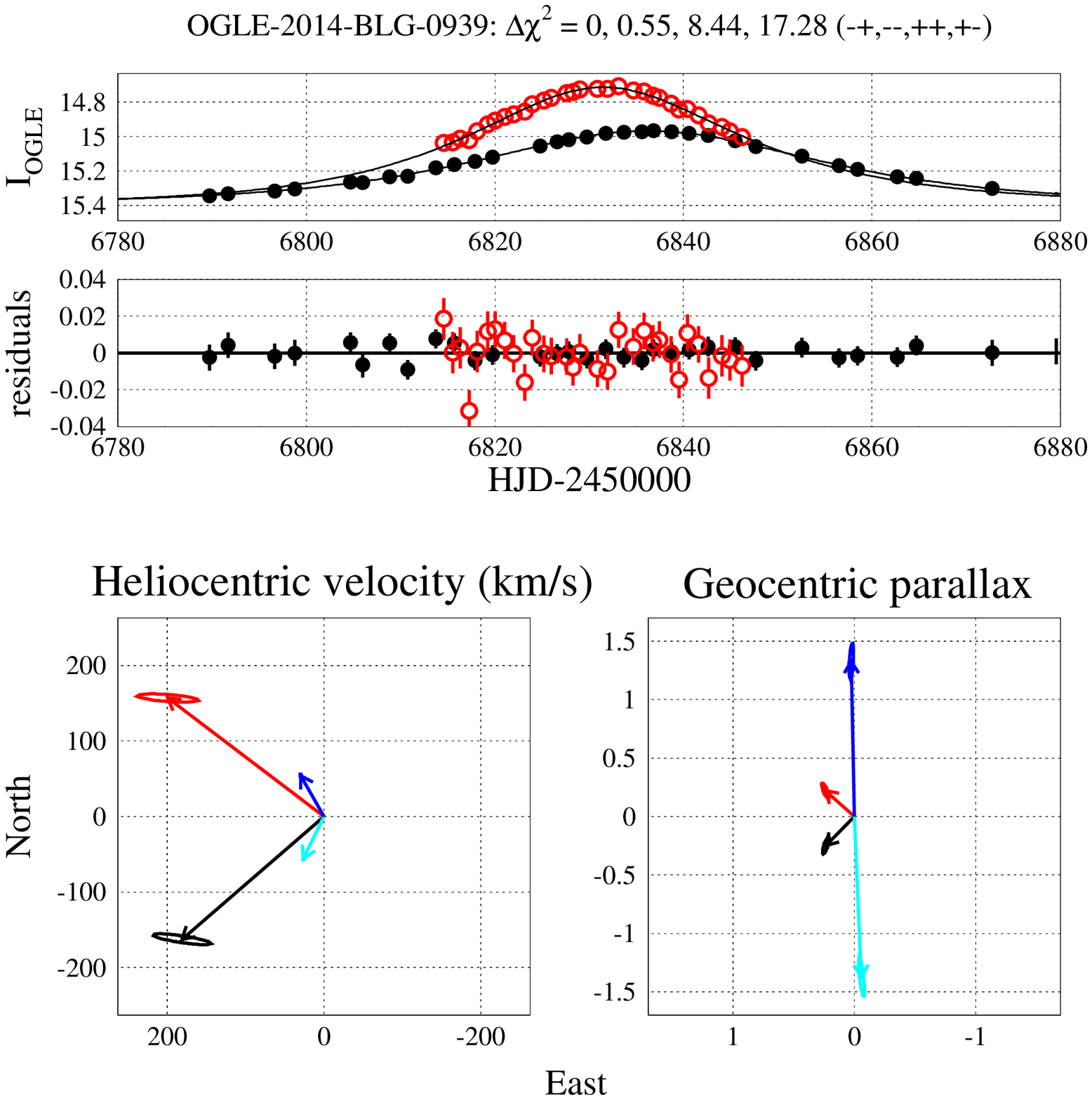}
\caption{OGLE-2014-BLG-0939. Panels and symbols as in Figure~\ref{fig:lc}a.}
\end{figure}
\clearpage

\begin{figure}
\plotone{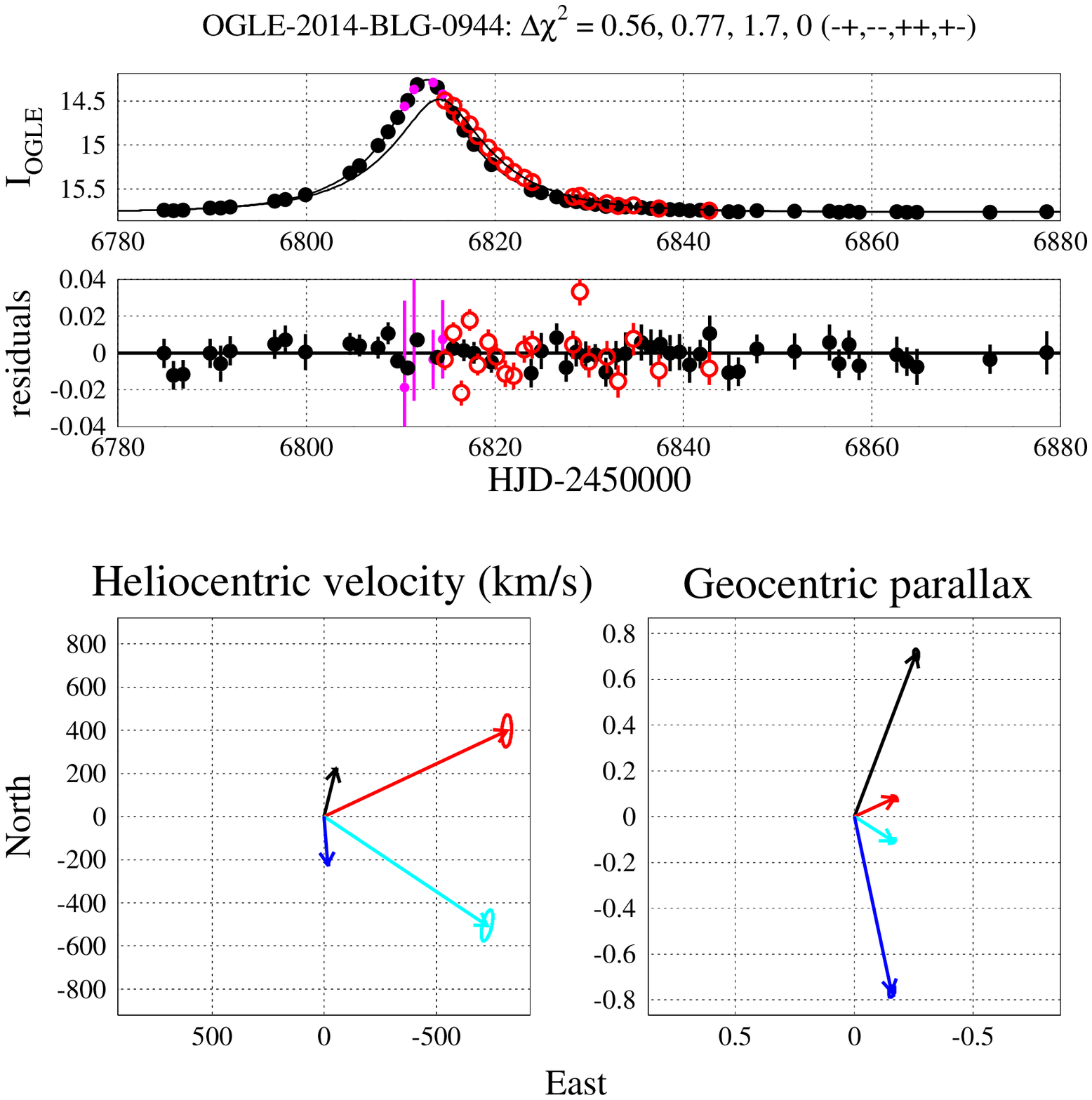}
\caption{OGLE-2014-BLG-0944. Panels and symbols as in Figure~\ref{fig:lc}a.}
\end{figure}
\clearpage

\begin{figure}
\plotone{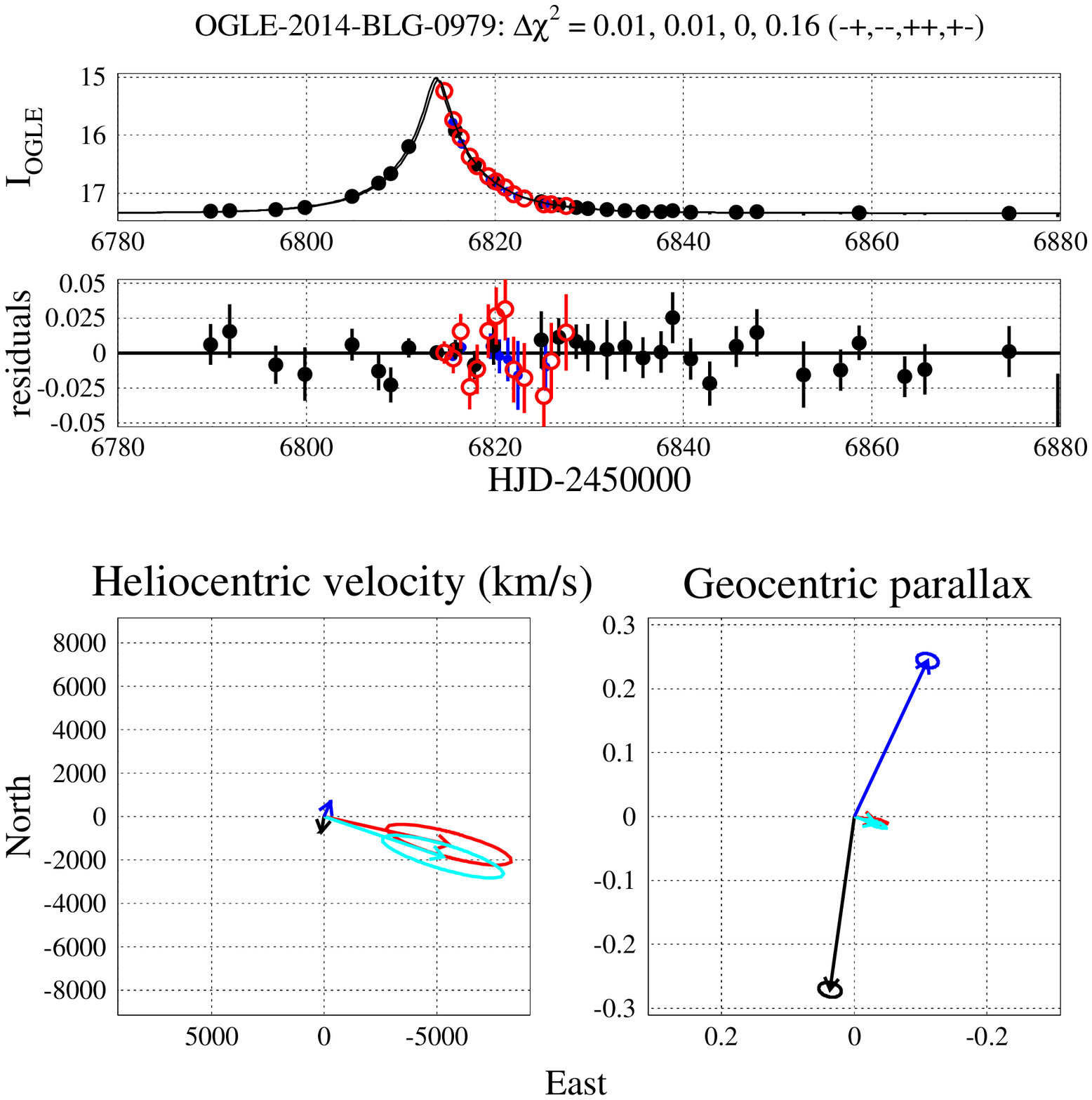}
\caption{OGLE-2014-BLG-0979. Panels and symbols as in Figure~\ref{fig:lc}a.}
\end{figure}
\clearpage

\begin{figure}
\plotone{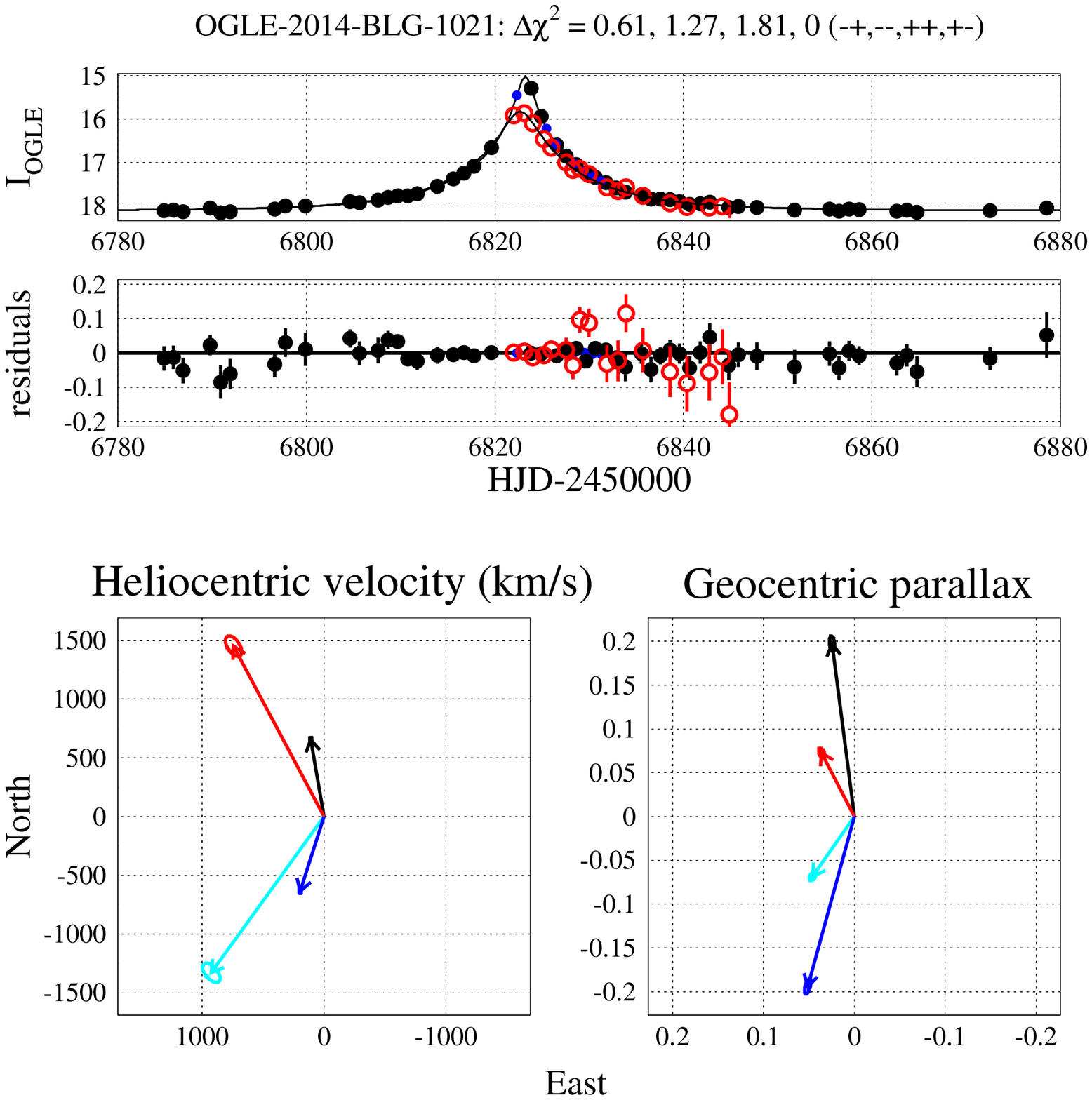}
\caption{OGLE-2014-BLG-1021. Panels and symbols as in Figure~\ref{fig:lc}a.}
\end{figure}
\clearpage

\begin{figure}
\plotone{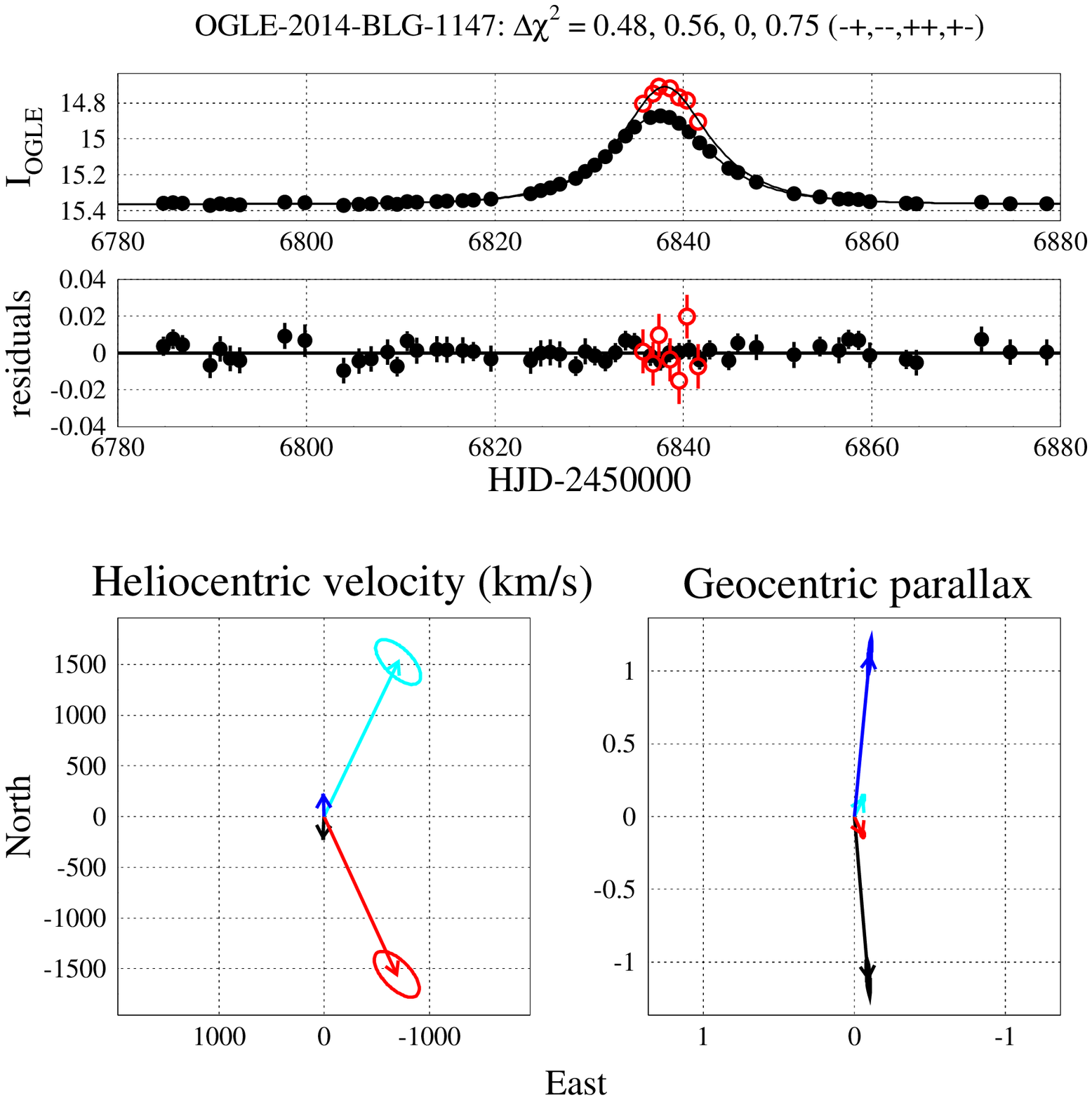}
\caption{OGLE-2014-BLG-1147. Panels and symbols as in Figure~\ref{fig:lc}a.}
\end{figure}

\end{subfigures}


\clearpage
\begin{figure}
\epsscale{.90}
\plotone{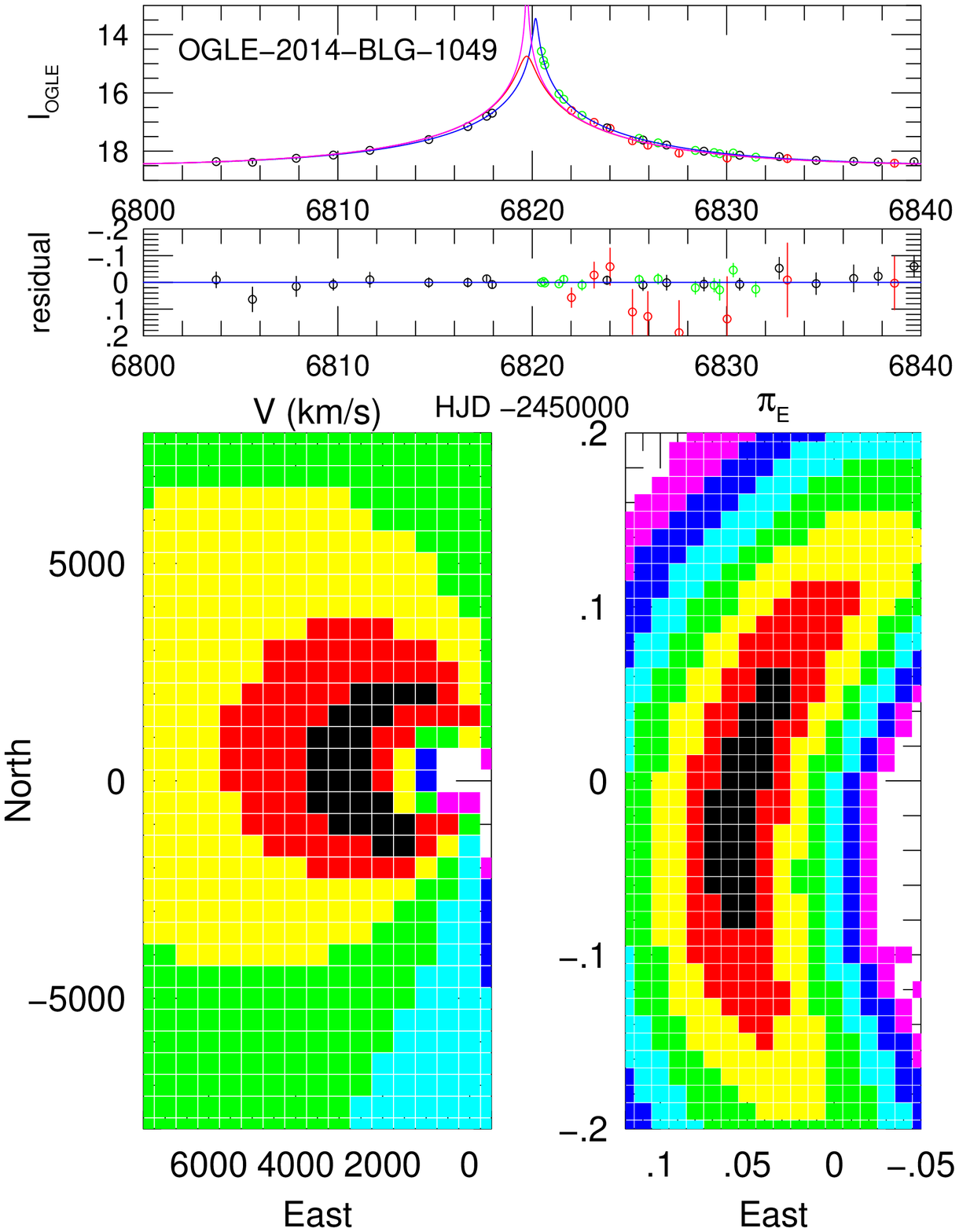}
\caption{Top Panel: OGLE-2014-BLG-1049 
lightcurves for OGLE (black), PLANET SAAO (green), and
{\it Spitzer} (red) data.  Ground-based model (blue) is well-defined
but many models are consistent with {\it Spitzer} data (e.g., red and
magenta curves).  Middle panel: residuals.  Lower panel: $\Delta\chi^2$
offsets (1, 4, 9, $\ldots$) from minimum for geocentric parallax $\bpi_{\e,\geo}$
(right) and heliocentric proper motion $\tilde\bv_\hel$ (left) for 
$u_{0,\oplus}>0$ solution (merger of $\Delta u_{0,\pm,+}$ solutions).
The $\Delta u_{0,\pm,-}$ solutions (not shown) are extremely similar.
Because $u_{0,Spitzer}$ is more poorly
defined that $t_{0,Spitzer}$ (top panel), $\Delta u_0$ is relatively
uncertain, which translates directly into uncertainty in $\pi_{\e,\rm north}$
because the Earth-{\it Spitzer} axis is almost due East-West.
}
\label{fig:1049}
\end{figure}

\begin{figure}
\epsscale{.80}
\plotone{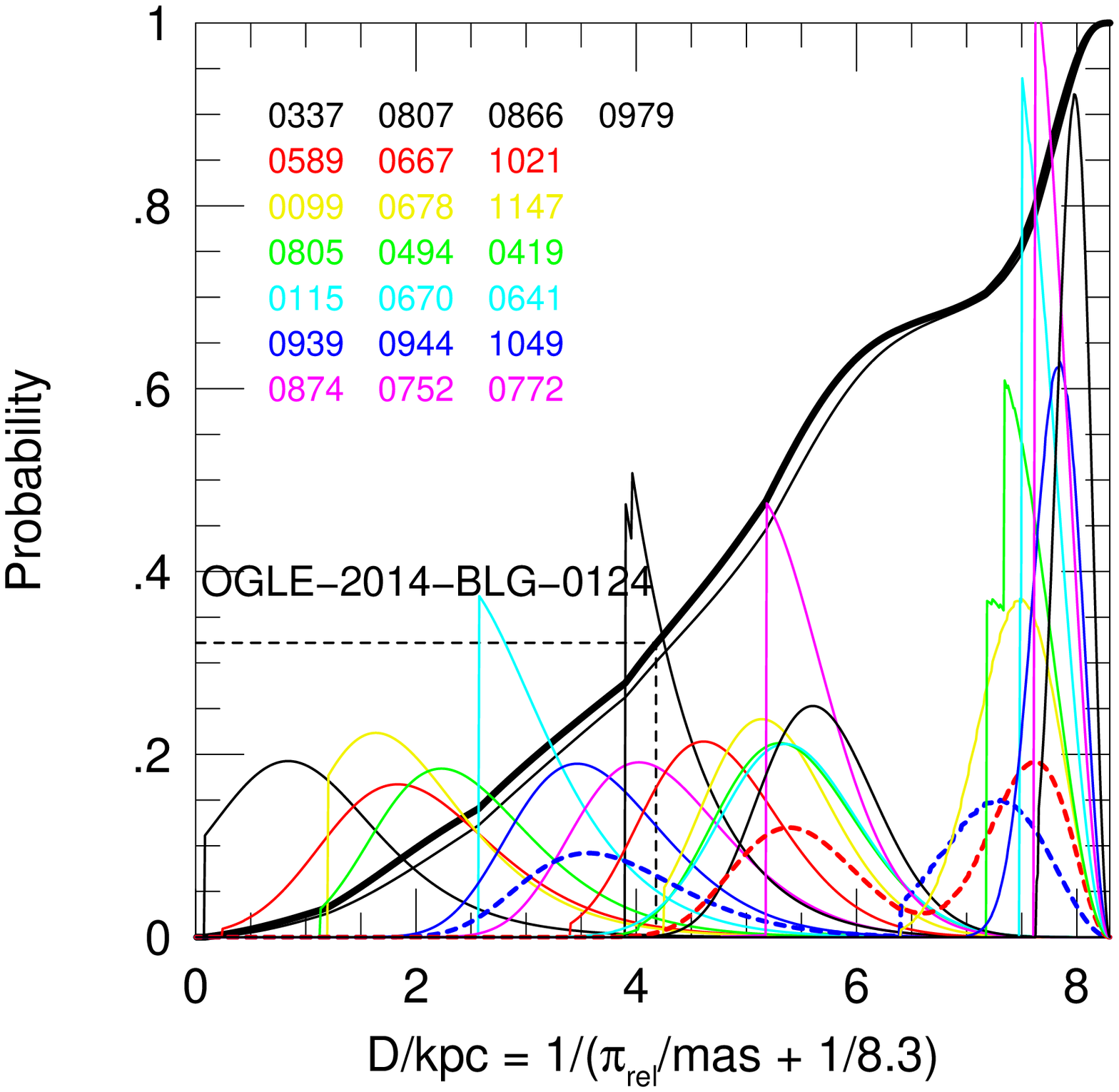}
\caption{Distance estimates for each of the 21 lenses analyzed in the
present paper plus OGLE-2014-BLG-0939 previously analyzed by \citet{ob140939}.
The curves indicate the individual probability distributions.  
The corresponding event names are listed
(upper left) in order of increasing mean estimated distance and
are displayed in the same color as the distribution.
The abscissa represents $D\equiv \kpc/(\pi_\rel/{\rm mas} + 1/8.3)$ because
it is the lens-source relative parallax $\pi_\rel$ that is actually measured.
With this display $D_L\sim D$ for $D\la D_S/2$ and $D_S-D_L \sim 8.3\,\kpc -D$
for $D\ga D_S/2$.  That is, the distance to the left boundary is very nearly
the lens distance for the left half of the diagram and the distance
to the right boundary is very nearly the distance between the lens and
source for the right half.  The value of $D$ for the one planet detected
by {\it Spitzer} in this campaign 
(orbiting the {\it lens} star in the event OGLE-2014-BLG-0124, 
\citealt{ob140124})
is shown by a dashed line.  By merging the results of several such campaigns
one would measure the Galactic distribution of planets between the
Solar circle and the Galactic bulge.  The calculation assumes a prior that
is flat in log-mass with hard cutoffs at
$M<1.1\,M_\odot$ (bulge) and $M<1.5\,M_\odot$ (disk).  
The cumulative distribution is shown for this calculation (bold) and
also for one with a realistic mass prior (solid).  The difference is
extremely small because the kinematic priors completely dominate.
Two events shown 
in bold dashed curves (OGLE-2014-BLG-0944 and OGLE-2014-BLG-1021) are the
only ones with ambiguous (disk/bulge) distance determinations.
}
\label{fig:cum}
\end{figure}

\end{document}